\documentclass[12pt]{article}

\usepackage[T1]{fontenc}
\usepackage[utf8]{inputenc}

\usepackage{amsmath,amsfonts,amssymb,amsthm,mathtools,dsfont}

\usepackage[semicolon,round,longnamesfirst]{natbib}

\usepackage[kerning=true,protrusion=true,expansion]{microtype}
\usepackage[a4paper,left=25mm,right=25mm,bottom=30mm,top=30mm]{geometry}
\usepackage{setspace}
\usepackage{comment}

\usepackage{sectsty}
\usepackage{titlesec}
\usepackage[title,titletoc,toc,page]{appendix}

\usepackage[usenames,dvipsnames]{xcolor}
\definecolor{mBlue}{HTML}{002FA7}
\definecolor{mGreen}{HTML}{009B55}
\definecolor{mOrange}{HTML}{FF4F00}

\usepackage{pgfplots,tikz,tikz-3dplot}
\pgfplotsset{compat=newest}
\usepgfplotslibrary{fillbetween}
\pgfplotsset{soldot/.style={color=blue,only marks,mark=*}}
\usetikzlibrary{positioning,decorations.markings,arrows,arrows.meta}
\usepgfplotslibrary{patchplots,ternary}

\usepackage{graphics,caption,subcaption}

\usepackage{enumerate}
\usepackage{accents}
\usepackage{url}
\usepackage{breakcites}
\usepackage[backref=page,breaklinks, pdftex]{hyperref}
\usepackage[nameinlink,noabbrev,capitalise]{cleveref}

\let\originalparagraph\paragraph
\renewcommand{\paragraph}[2][.]{\originalparagraph{#2#1}} 

\paragraphfont{\color{mGreen!50!black}}

\hypersetup{
	backref=true, 
	pagebackref=true, 
	hyperindex=true, 
	colorlinks=true, 
	breaklinks=true, 
	urlcolor=mBlue, 
	linkcolor=mBlue,  
	citecolor=NavyBlue 
	}

\theoremstyle{plain}
\newtheorem{lemma}{Lemma}
\newtheorem{proposition}{Proposition}
\newtheorem{theorem}{Theorem}
\newtheorem*{theorem*}{Theorem}
\newtheorem{corollary}{Corollary}

\newtheorem{definition}{Definition}

\newcommand{\ubar}[1]{\underaccent{\bar}{#1}}

\makeatletter
\def\mathcolor#1#{\@mathcolor{#1}}
\def\@mathcolor#1#2#3{%
  \protect\leavevmode
  \begingroup
    \color#1{#2}#3%
  \endgroup
}
\makeatother

\title{\textbf{Persuading a Wishful Thinker}\thanks{We especially thank Jeanne Hagenbach and Eduardo Perez-Richet for their comments and support. We also thank S. Nageeb Ali, Sarah Auster, Gregorio Curello, Roland Bénabou, Geoffroy de Clippel, Alexis Ghersengorin, Simon Gleyze, Emeric Henry, Deniz Kattwinkel, Frédéric Koessler, Laurent Mathevet, Meg Meyer, Daniel Monte, Franz Ostrizek, Nikhil Vellodi, Adrien Vigier and Yves Le Yaouanq as well as seminar audiences at Sciences Po, Paris School of Economics, S\~ao Paulo School of Economics (FGV) and at the EEA-ESEM 2021 for valuable feedback and comments at various stages of the project. Part of this research was carried out while both authors were at the Department of Economics of Sciences Po. This project has received funding from the European Research Council (ERC) under the European Union's Horizon 2020 research and innovation programme (grant agreement n°850996 -- MOREV and n°101001694 -- IMEDMC). All remaining errors are ours.}}
\author{Victor Augias\thanks{University of Bonn, Department of Economics, e-mail: \href{mailto:vaugias@uni-bonn.de}{\texttt{vaugias@uni-bonn.de}}.} \and Daniel M. A. Barreto\thanks{University of Amsterdam, e-mail: \href{mailto:d.m.a.barreto@uva.nl}{\texttt{d.m.a.barreto@uva.nl}}.}}
\date{\today}

\begin{document}

\maketitle

\begin{abstract}
    We study a persuasion problem in which a sender designs an information structure to induce a non-Bayesian receiver to take a particular action. The receiver, who is privately informed about his preferences, is a wishful thinker: he is systematically overoptimistic about the most favorable outcomes. We show that wishful thinking can lead to a qualitative shift in the structure of optimal persuasion compared to the Bayesian case, whenever the sender is uncertain about what the receiver perceives as the best-case outcome in his decision problem.
\end{abstract}

\section{Introduction}

Information is a powerful tool for shaping behavior. But unlike material incentives, information is distinctive in that the way and the extent to which it influences individuals' behavior is highly sensitive to the way in which they process it. While most of the insights derived in the information design literature rely on the assumption that decision-makers on the receiving end of information are Bayesian, and thus able to optimally infer the informational content of any realized signal, a substantial literature in psychology and behavioral economics shows that the process by which individuals interpret information and form beliefs is not guided solely by the search for accuracy, but often depends on their wishes and desires. An usual manifestation of this is \emph{wishful thinking}: the tendency for individuals to let their \emph{preferences about outcomes} influence the way they process information, leading to beliefs that are systematically biased towards outcomes they wish
to be true.\footnote{There exists abundant empirical and experimental evidence of wishful thinking. See in particular \cite{Benabou2016}, page 150 and \cite{Benjamin2019} Section 9, as well as, e.g., \cite{Weinstein1980}, \cite{Mijovic-Prelec2010}, \cite{Mayraz2011a}, \cite{Heger2018}, \cite{Coutts2019},
\cite{Engelmann2019} or \cite{Jiao2020}.} 

In this paper we investigate how wishful thinking affects the nature of optimal information design. We do so by studying a setting in which a sender designs an information structure with the goal of inducing a wishful receiver, who is privately informed about his own preferences, to take a particular action. Receiver's wishful belief updating is modeled as in \cite{Caplin2019}: after observing a signal realization, receiver forms a wishful belief by trading-off the anticipated payoff associated with that belief against the cost of distorting his belief away from the Bayesian one. As wishful belief updating ties the way the receiver processes information with his preferences, any uncertainty the sender might have about receiver's preferences translates into \emph{uncertainty about how the receiver processes information}.

Our main result, \cref{thm:opt_info}, shows that wishful thinking can \emph{reverse} the structure of optimal persuasion depending on the type of uncertainty the sender has about the receiver's preferences. We show that in persuasion environments where the Bayesian-optimal information structure would exhibit \emph{upper-censorship} \citep{Kolotilin2022}, the wishful-optimal information structure might instead exhibit \emph{lower-censorship} if the sender is uncertain about what the receiver perceives as the best-case outcome in his decision problem.

To prove this result, we first characterize how wishful thinking skews the receiver's behavior relative to the behavior of a Bayesian receiver. \cref{thm:behavior_equivalence} establishes the equivalence between the optimal behavior of a wishful receiver and the behavior of a ``virtual'' Bayesian receiver whose material payoff is given by an exponential transformation of the receiver's original payoff function. We say that wishful thinking leads the receiver to \emph{motivate} some action whenever such virtual receiver takes such action under a strictly larger set of Bayesian posteriors than a Bayesian receiver endowed with the original payoff function would. Our \cref{thm:motivatedaction} then characterizes, for any possible binary decision problem, which action is motivated by a wishful receiver as a function of his material payoff and of the magnitude of his belief distortion cost: agents with a low cost of belief distortion motivate the action associated with the highest possible payoff in the decision problem, whereas agents with a high cost of belief distortion motivate the action associated with the highest payoff variability across states. In our model, receiver's private type defines both their \emph{directional motive}, that is, which of the two actions they motivate, and the \emph{intensity} of their behavioral distortion, that is, the size of the set of Bayesian posteriors for which they behave differently than a Bayesian receiver would.

Knowing the receiver's directional motive is relevant for the sender's optimal information design, as it defines the \emph{type and quantity of information} that the receiver requires in order to choose sender's preferred action. Whenever the sender knows receiver's directional motive, we show that the wishful-optimal information structure is qualitatively identical to the Bayesian-optimal one, generating one signal that induces the receiver to reject the sender's preferred action for sure and one signal that induces the receiver to opt for sender's preferred action with high enough probability. Wishful-optimal information structures in this case only differ from Bayesian-optimal ones in their informativeness. This is not the case if the sender is uncertain about receiver's directional motive. As in this case the sender is uncertain whether receiver's wishfulness causes him to require more or less information in order to adopt the intended action, the optimal information structure hedges against sender's uncertainty by only generating signals that induce the intended action with positive probability: it generates one conclusive signal that leads every receiver type to adopt sender's preferred action, and one inconclusive signal that still convinces some receiver types (those whose distortion is in line with sender's interests) to act as intended by the sender.

\paragraph{Related literature}

The literature on persuasion and information design, initiated by \citeauthor{Kamenica2011}'s (\citeyear{Kamenica2011}) seminal contribution, has focused on characterizing how and to what extent a sender with full commitment power can influence the behavior of \emph{Bayesian} decision-makers by strategically disclosing information.\footnote{See \cite{Bergemann2019} and \cite{Kamenica2019} for thorough reviews of this literature.} By considering a receiver prone to wishful thinking, we contribute to the literature that studies the persuasion of receivers making \emph{non-Bayesian inferences}.\footnote{See \cite{Benjamin2019} for an in-depth review of the literature exploring how statistical inferences made by individuals deviate from Bayes' rule. In particular, wishful thinking belongs to the family of preference-based inferences examined in Section 9.}

An important contribution in that respect is that of \cite{DeClippel2022}. The authors identify a substantive class of non-Bayesian updating rules under which there exists a belief distortion function that maps any Bayesian posterior belief to the (unique) receiver’s subjective belief. They use their characterization to show that the concavification approach of \cite{Kamenica2011} carries over in many non-Bayesian persuasion problems considered in the literature.\footnote{See also \cite{Aumann1995} for an earlier use of the concavification for characterizing the equilibria of repeated games with incomplete information.} The receiver in our model has a belief updating rule that falls into the class identified by \cite{DeClippel2022}, so that we can also successfully apply the concavification approach to persuading a wishful receiver. Vis-à-vis \cite{DeClippel2022}, our contribution lies in the complete characterization of how the receiver's wishful belief updating rule, introduced by \cite{Caplin2019}, affects the optimal choice of information structure for the sender. Other contributions analyze how specific non-Bayesian updating rules affect optimal persuasion. \cite{Levy2022} analyze a non-Bayesian persuasion setting where the receiver is subject to correlation neglect. \cite{Galperti2019} analyzes a model of persuasion where the receiver updates beliefs in a non-Bayesian way after observing unexpected news, while \cite{Eliaz2021,Eliaz2021a} analyze persuasion models where the receiver's belief updating rule is misspecified. Our work also relates to other models examining the effects of the receiver’s non-Bayesian updating on optimal information disclosure, but considering communication protocols that do not make the full commitment assumption. \cite{Lee2023} investigate the implications of prior-biased inferences in cheap-talk, while \cite{Benjamin2019a} analyze an example where a sender discloses verifiable information to an audience inclined to base-rate neglect. In contrast, we study how wishful thinking affects persuasion when the sender has full commitment power over the information structure.

Our paper is also part of a literature that studies how information should be disclosed to ``psychological'' receivers, i.e., receivers who intrinsically value their beliefs. The contributions closest to ours in this regard are that of \cite{Lipnowski2018}, who investigate how a sender with full commitment power should design information to persuade a psychological receiver, and \cite{Hagenbach2022}, who analyze how a sender should disclose verifiable information to a psychological receiver. Although beliefs also affect the receiver's utility in these models, he is nonetheless perfectly Bayesian. The belief-based utilities thus only influence how the receiver values information but do not determine his belief updating. Instead, we study optimal information design when the receiver's posterior belief is determined by the maximization of his belief-based psychological utility.

Our model also shares similarities with persuasion models where the receiver is rationally inattentive as in \cite{Bloedel2021}, \cite{Lipnowski2020,Lipnowski2022} and \cite{Wei2021}. In this series of papers, the sender seeks to persuade a receiver for whom processing information is costly. The similarity lies in the form of the receiver's attention cost function---defined as the expected reduction in the receiver's belief entropy---with the receiver's belief distortion function in our model---defined as the Kullback-Leibler divergence between the receiver's subjective belief and the Bayesian posterior belief. The fundamental distinction between these models and ours is that the attention cost function is defined over the set of information structures and not over the set of beliefs: the inattentive receiver chooses a subjective information structure given the information structure designed by the sender, whereas a wishful receiver chooses a subjective posterior belief given the Bayesian posterior induced by sender's information policy. Furthermore, although inattentive, the receiver interprets the information resulting from his attention decision in a Bayesian way.

Methodologically, our work is related to two recent strands of work in Bayesian persuasion. Firstly, although our baseline physical environment corresponds to the \emph{linear} persuasion framework, thanks to which most of the meaningful progresses have been achieved in the literature \cite[see][among others]{Kolotilin2017,Kolotilin2018,Dworczak2019,Guo2019,Dizdar2020,Kleiner2021,Kolotilin2022,Arieli2023,Candogan2023}, our \Cref{thm:behavior_equivalence} establishes an equivalence between our non-Bayesian linear persuasion problem and a Bayesian persuasion problem whose receiver has \emph{nonlinear preferences}. Solving non-linear persuasion problems has proven to be challenging and requires either making assumptions about the shape of the receiver's payoff function that are typically not satisfied in our model (see in particular \citealp{Kolotilin2023} and \citealp{Smolin2023}) or working at such a high level of generality that only qualitative properties about the structure of optimal solutions can be obtained \citep{Dworczak2023}. Our contribution therefore constitutes a modest step towards the characterization of optimal information structures in yet unsolved non-linear persuasion problems. Secondly, a recent body of work has focused on studying how the nature and informativeness of the sender's optimal policy are affected by a shift in the primitives of the persuasion model. Such a comparative statics exercise has been initiated by \cite{Kolotilin2015}, extended by \cite{Kolotilin2022}, and now culminates in the analysis of \cite{Curello2023} who characterize, in the linear persuasion model, which type of shifts in the sender's indirect utility function guarantee that the optimal information policy of the sender is more or less informative. The contribution that is the most related to our paper in that strand of the literature is \cite{Gitmez2023}. The authors analyze a binary linear persuasion model in which an informational autocrat wants to persuade population of citizens, that are both heterogeneous in their preferences and prior beliefs, to support a policy.\footnote{Although we frame our model as having a single receiver, interpreting it in terms of a continuous unit mass of agents is also valid.} They compare the nature of the optimal persuasion policy as a function of the degree of dispersion of citizens' opinions, captured by their preferences and priors. Persuading a society whose distribution of opinions is not very dispersed---modeled by the fact that the density of the distribution of opinions is single-peaked---entails using an upper-censorship policy, whereas persuading a society whose distribution of opinions is very dispersed---modeled by the fact that the density of the distribution of opinions is single-dipped---entails using a lower-censorship policy. Although our main result, \cref{thm:opt_info}, shows a similar qualitative change in the structure of the sender's optimal information policy, the origin of this change is the exogenous variation in the distribution of opinions in their model, whereas it is endogenously generated by the receiver's wishful thinking in ours.

\section{Model}

\subsection{Setting}

\paragraph{Actions, states and material payoffs}

The model involves two agents: a sender (referred to as `she') and a receiver (referred to as `he'). The underlying state of the world is denoted by $\omega\in \Omega=\{0,1\}$. Both the sender and the receiver possess a common prior belief, $\mu_{0}\in\Delta(\Omega)$, regarding the state.\footnote{For any set $X$ the set $\Delta(X)$ denotes the space of all Borel probability measures over $X$.}

The receiver selects an action $a$ from the set $A = \{0,1\}$ and is privately informed of his type, which is given by the tuple  $(\gamma,\theta)\in T=\{0,1\}\times\mathopen[0,1\mathclose]$. We assume that $\gamma$ and $\theta$ are independently distributed random variables and we let $p\in\mathopen[0,1\mathclose]$ be the probability that $\gamma=1$ and $F\colon\Theta\to\mathopen[0,1\mathclose]$ be the cumulative distribution function of $\theta$. We assume that $F$ admits a density function $f\colon \Theta \to \mathbb{R}$ that is strictly positive, continuously-differentiable and log-concave.\footnote{The family of log-concave densities is a broad class of single-peaked probability densities. For a comprehensive list of log-concave distributions, we refer to \cite{Bagnoli2005}.}

The payoff for a receiver of type $(\gamma,\theta)$, upon selecting action $a$ in state $\omega$, is given by
\begin{equation*}
    u(a,\omega,\gamma,\theta) = a(\omega-\theta) + \gamma \theta (1-\omega),
\end{equation*}
This specification corresponds to a slightly more general version of the usual linear payoff function for the receiver that is considered in the literature \cite[see, e.g.,][]{Kolotilin2017}. We defer the explanation for the significance and role of the type $(\gamma,\theta)$ to \cref{sec:implications} as it becomes clearer once \cref{thm:implications_linear} is stated.

The sender's objective is to persuade the receiver to opt for action $a=1$, irrespective of the actual state. The sender's payoff, when the receiver selects action $a$, is given by
\begin{equation*}
    v(a) = a,
\end{equation*}
for any $a\in A$.\footnote{Our main result would remain unchanged if we considered a partially aligned sender whose preferences are given by
\begin{equation*}
    v(a,\omega,\gamma,\theta)=a+\lambda u(a,\omega,\gamma,\theta),
\end{equation*}
as long as the the weight $\lambda>0$ the sender assigns to the welfare of the receiver is not too large, in which case full information disclosure would be optimal.}

\paragraph{Information and sender's beliefs}

The sender influences the receiver’s behavior by designing an information structure, denoted as $(\sigma,S)$ which includes an exogenous and sufficiently rich set of signal realizations, $S$, and a communication strategy, $\sigma\colon\Omega\to\Delta(S)$. This strategy maps any given state, $\omega$, to a conditional distribution, $\sigma(\omega)$, over the possible signal realizations.

When the sender observes a particular signal realization, $s\in S$, she updates her prior belief using Bayes' rule. Her posterior belief after observing a signal realization $s$ can therefore be expressed as:
\begin{equation*}
     \mu(\omega \, | \, s) = \frac{\sigma(s \, | \, \omega) \, \mu_{0}(\omega)}{\displaystyle\sum_{\omega\in\Omega}\sigma(s \, | \, \omega) \, \mu_{0}(\omega)},
\end{equation*}
for every $\omega\in \Omega$.

\paragraph{Receivers' beliefs and behavior}

In contrast with the sender, the receiver is a wishful thinker. His subjective belief trades-off the psychological benefit, in terms of anticipated utility, of deluding oneself against the cost of distorting away from the Bayesian belief. Formally, his \emph{anticipated material payoff} under some action $a$ given his subjective belief $\eta$ is given by
\begin{equation*}
    U(a,\eta,\gamma,\theta) = \sum_{\omega\in\Omega} u(a,\omega,\gamma,\theta)\, \eta(\omega),
\end{equation*}
for any $a\in A$, $\eta\in\Delta(\Omega)$, and $(\gamma,\theta)\in T$, while his \emph{belief distortion cost} is given by the Kullback-Leibler divergence between his subjective belief $\eta$ and the Bayesian posterior $\mu$ induced by any signal realization, defined by
\begin{equation*}
    C(\eta,\mu) = \sum_{\omega\in\Omega} \eta(\omega)\ln\left(\frac{\eta(\omega)}{\mu(\omega)}\right),
\end{equation*} 
Actions and beliefs of the receiver are then jointly determined by the following optimization program\footnote{As already noted by \cite{Bracha2012} as well as  \cite{Caplin2019}, the objective function in problem \labelcref{eq:receiverproblem} has a similar mathematical structure to the multiplier preferences introduced by \cite{Hansen2001} and axiomatized by \cite{Strzalecki2011}. Precisely, the decision-maker in \cite{Strzalecki2011} solves
\begin{equation*}
    \max_{a\in A}\min_{\eta\in\Delta(\Omega)}\sum_{\omega\in\Omega} u(a,\omega)\,\eta(\mathrm{d}\omega)+\frac{1}{\rho}C(\eta,\mu),
\end{equation*}
for any given $\mu\in\Delta(\Omega)$. In that model, the parameter $\rho$ measures the degree of confidence of the decision-maker in the belief $\mu$ or, in other words, the importance he attaches to belief misspecification. Conclusions regarding the belief distortion in that setting would be reversed with respect to our model, as a receiver forming beliefs according to that non-Bayesian kind of updating would form overpessimistic beliefs compared to the sender's Bayesian posterior belief.}:
\begin{equation}\label{eq:receiverproblem}
    \max_{a\in A} \max_{\eta\in\Delta(\Omega)} \; U(a,\eta,\gamma,\theta) - \frac{1}{\rho} C(\eta,\mu),
\end{equation}
where $\rho\in\mathopen(0,+\infty\mathclose)$ is a parameter scaling the distortion cost, which we interpret as the receiver's degree of \emph{wishfulness}.\footnote{We should also mention that all our result would hold verbatim if $\rho$ corresponded to an increase in the receiver's \emph{material stakes}. Indeed, problem \labelcref{eq:receiverproblem} has the same solution than one whose material payoff function would have been inflated by a factor $\rho$:
\begin{equation*}
    \max_{a\in A}\min_{\eta\in\Delta(\Omega)}\sum_{\omega\in\Omega} \rho u(a,\omega)\,\eta(\mathrm{d}\omega)-C(\eta,\mu).
\end{equation*}
}

We denote by $A(\mu, \rho, \gamma, \theta)$ the set of optimal actions for a receiver of wishfulness degree $\rho$ and type $(\gamma,\theta)$ whenever the realised Bayesian posterior is $\mu$, defined by
\begin{equation}\label{eqn:wishful_actions}
    A(\mu, \rho, \gamma, \theta)=\underset{a\in A}{\arg\max} \max_{\eta\in\Delta(\Omega)} \; U(a,\eta,\gamma,\theta) - \frac{1}{\rho} C(\eta,\mu),
\end{equation}
for every $\mu\in\Delta(\Omega)$, $\rho\in\mathopen(0,+\infty\mathclose)$ and $(\gamma,\theta)\in T$.

\paragraph{Information design problem}

We slightly abuse notation by henceforth letting $\mu\in \mathopen[0,1\mathclose]$ stand for the probability that the sender assigns to state $\omega=1$ under his posterior belief. We know from \cite{Kamenica2011}, Proposition 1, that we can equivalently think of Sender committing ex-ante to an information structure $(\sigma, S)$ or to an \emph{information policy} $\pi\in \Pi(\mu_0)$, where
\begin{equation*}
  \Pi(\mu_0)=\left\{\pi\in\Delta\big(\mathopen[0,1\mathclose]\big) \, \Big| \, \int_{\mathopen[0,1\mathclose]}\mu \, \pi(\mathrm{d}\mu)=\mu_0 \right\}.
\end{equation*}
As is usual in the persuasion literature, we focus on sender-preferred equilibria. We let $a(\mu,\rho,\gamma,\theta)\in A(\mu,\rho,\gamma,\theta)$ denote the sender's preferred selection in $A(\mu,\rho,\gamma,\theta)$. Since the sender does not know the receiver's types $(\gamma,\theta)$, the relevant statistic for the sender is the expected action of the receiver at any posterior belief $\mu$, given by
\begin{equation*}
    V(\mu,\rho) = \int_{0}^{1} \bigl(p \,  a(\mu,\rho,1,\theta) + (1-p) a(\mu,\rho,0,\theta)\bigr) \, f(\theta) \, \mathrm{d}\theta ,
\end{equation*}
for all $\mu\in\mathopen[0,1\mathclose]$ and $\rho\in\mathopen(0,+\infty\mathclose)$. The information design problem of the sender consists in choosing the information policy maximizing the ex-ante probability that the receiver takes action $a=1$:
\begin{equation*}
    \max_{\pi\in\Pi(\mu_{0})} \; \int_{\mathopen[0,1\mathclose]} V(\mu,\rho) \, \pi(\mathrm{d}\mu).
\end{equation*}

\subsection{Discussion of the belief distortion cost function}

Wishful thinking is one possible manifestation of \emph{motivated reasoning} \citep{Kunda1987,Kunda1990}. It has been shown empirically and experimentally that motivated reasoning leads individuals to use sophisticated mental strategies such as manipulating their own memory\footnote{See \cite{Benabou2015} and \cite{Benabou2016}. For experimental evidence on memory manipulation see, e.g., \cite{Saucet2019}, \cite{Carlson2020} and \cite{Chew2019}.}, avoiding freely available information\footnote{See \cite{Oster2013} for compelling empirical evidence, and \cite{Golman2017} and references therein} or creating elaborate narratives supporting their bad choices or inaccurate claims to justify their preferred beliefs.\footnote{One can relate this possible microfoundation of the belief distortion cost to the literature on lying costs \citep{Abeler2014,Abeler2019} since, when Receiver is distorting away his subjective belief from the rational Bayesian beliefs, he is essentially lying to himself. We thank Emeric Henry for suggesting us this interpretation of the cost function.} Our assumption that the cost function is the Kullback-Leibler divergence between the receiver's subjective belief and the Bayesian posterior captures, in ``reduced form'', the fact that implementing such mental strategies comes at a cost when desired beliefs diverge from what evidence suggests.

Another type of cost function considered in the literature was introduced by \cite{Brunnermeier2005}. In contrast to the Kullback-Leibler cost function, which captures psychological cost of belief distortion, \citeauthor{Brunnermeier2005}'s (\citeyear{Brunnermeier2005}) cost is the expected \emph{material loss} induced by inaccurate beliefs, given by
\begin{equation*}
    C^{BP}(\eta,\mu)=\mathbb{E}_{\mu}\left[\max_{a\in A(\mu)} u(a,\omega)-\max_{a\in A(\eta)} u(a,\omega)\right],
\end{equation*}
for any Bayesian posterior $\mu\in\Delta(\Omega)$ and subjective belief choice $\eta\in\Delta(\Omega)$, where $u(a,\omega)$ is the material payoff function of the agent and $A(\mu)$ is the optimal choice correspondence of the decision-maker at belief $\mu\in\Delta(\Omega)$.\footnote{This function has also been used by \cite{Gossner2018} to analyze the cost of misperceptions and by \cite{Frankel2019} to define the value of information.}


By using the Kullback-Leibler specification for the cost function, we follow the earlier attempts by \cite{Bracha2012} and \cite{Caplin2019} to model motivated reasoning as an optimal choice of belief. The motivation behind this modelling assumption is twofold. Firstly, \cite{Coutts2019} provides a an experimental test determining if and which of the \citeauthor{Bracha2012}'s/\citeauthor{Caplin2019}'s and \citeauthor{Brunnermeier2005}'s specifications for the belief distortion cost function are in line with how subjects distort their beliefs when they are incentivized to be overoptimistic. Although the author provides mixed evidence, showing that subjects' beliefs significantly differ from what Bayes' rule would predict, but that none of the costs fully explain this distortion, he does show that raising the stakes for accuracy leads subjects' beliefs to deviate from Bayesian ones in a way which is consistent with the psychological cost of belief distortion. Secondly, working with the Kullback-Leibler divergence has the advantage, unlike the material cost function, of producing closed-form solutions for the receiver's optimal belief that are non-degenerate. In our binary setting, a wishful agent having a \citeauthor{Brunnermeier2005}'s type of cost would always fully delude himself by placing probability one on either of the states.

\section{Optimal persuasion of a wishful receiver}\label{sec:behavior}

\subsection{Behavioral implications of wishful thinking}

Before discussing how wishful thinking impacts persuasion, we must first understand how wishful thinking affects the behavior of the receiver. The results in this section answer this question. Those results do not require the linear specification for the receiver's payoff function. We further discuss what these results imply for the linear specification in \cref{sec:implications}.

Let $A = \{0,1\}$ and $\Omega = \{0,1\}$ be the action and state space and consider any payoff function $u\colon A \times \Omega \to \mathbb{R}$. Let also $A(\mu,\rho)$ be the set of wishful optimal actions, defined in the same way as in \labelcref{eqn:wishful_actions}:
\begin{equation*}
    A(\mu,\rho)=\underset{a\in A}{\arg\max} \max_{\eta\in\Delta(\Omega)} \; \sum_{\omega\in\Omega} u(a,\omega) \, \eta(\omega) - \frac{1}{\rho} \sum_{\omega\in \Omega} \eta(\omega) \ln\left(\frac{\eta(\omega)}{\mu(\omega)}\right),
\end{equation*}
for any $\mu\in\mathopen[0,1\mathclose]$ and $\rho\in\mathopen(0,+\infty\mathclose)$.
\begin{proposition}[\citealp{Caplin2019,Robson2023}]\label{thm:behavior_equivalence}
    The optimal action correspondence of a wishful receiver is given by
    \begin{equation*}
        A(\mu,\rho) = \underset{a\in A}{\arg\max} \sum_{\omega\in \Omega} e^{\rho \, u(a,\omega)} \, \mu(\omega),
    \end{equation*}
    for any $\mu\in\Delta(\Omega)$ and $\rho\in\mathopen(0,+\infty\mathclose)$.
\end{proposition}
The above result shows that, from the sender's point of view, the behavior of a wishful receiver is equivalent to that of a \emph{virtual} Bayesian receiver whose payoff function corresponds to an exponential distortion of the receiver's original material payoff. Although we state this proposition for binary action and state spaces for the sake of consistency, it applies more generally. For completeness, we provide in \cref{secap:behavior_equivalence_proof} a proof of \cref{thm:behavior_equivalence} together with a characterization of the optimal wishful beliefs in a more general case than in \cite{Caplin2019} and \cite{Robson2023}. Namely, when $A$ is arbitrary and $\Omega$ is a Polish space. Intuitively, this equivalence comes from the fact that
\begin{equation*}
   \frac{1}{\rho}\ln\left(\sum_{\omega\in \Omega} e^{\rho \, u(a,\omega)} \, \mu(\omega)\right) = \max_{\eta\in\Delta(\Omega)} \sum_{\omega\in\Omega} u(a,\omega) \, \eta(\omega) - \frac{1}{\rho} \sum_{\omega\in \Omega} \eta(\omega) \ln\left(\frac{\eta(\omega)}{\mu(\omega)}\right)
\end{equation*}
for every $a\in A$ and $\mu\in\Delta(\Omega)$. Which implies that a wishful receiver's optimal choice of action corresponds to the solution of the problem
\begin{equation*}
    \max_{a\in A} \frac{1}{\rho}\ln\left(\sum_{\omega\in \Omega} e^{\rho \, u(a,\omega)} \, \mu(\omega)\right),
\end{equation*}
which has the same set of maximizers than the problem
\begin{equation*}
    \max_{a\in A} \sum_{\omega\in \Omega} e^{\rho \, u(a,\omega)} \, \mu(\omega).
\end{equation*}

Coming back to the case with binary action and state spaces, the optimal behavior of a wishful receiver is thus pined-down by a single belief indifference cutoff $\mu_W(\rho)\in\mathopen[0,1\mathclose]$ given by
\begin{equation*}
    \mu_W(\rho) = \frac{e^{\rho u(0,0)} - e^{\rho u(1,0)}}{e^{\rho u(0,0)} - e^{\rho u(1,0)} + e^{\rho u(1,1)} - e^{\rho u(0,1)}}.
\end{equation*}
for any $\rho\in\mathopen(0,+\infty\mathclose)$, and such that:
\begin{equation*}
    A(\mu,\rho)=\left\{\begin{array}{ll}
        \{0\} & \text{if $\mu<\mu_W(\rho)$} \\
        \{0,1\} & \text{if $\mu=\mu_W(\rho)$} \\
        \{1\} & \text{if $\mu>\mu_W(\rho)$}
    \end{array}\right.,
\end{equation*}
for any $\mu\in\mathopen[0,1\mathclose]$. In comparison, the belief indifference cutoffs of a Bayesian receiver is given by
\begin{equation*}
        \mu_B = \frac{u(0,0)  - u(1,0)}{u(0,0)  - u(1,0) + u(1,1) - u(0,1)}.
\end{equation*}
Hence, a wishful receiver behaves differently than a Bayesian one at some subset of posterior beliefs whenever $\mu_B \neq \mu_W(\rho)$. Furthermore, whether the wishful cutoff is greater or smaller than the Bayesian cutoff defines the direction in which an agent's wishfulness skews his behavior.
\begin{definition}[Motivated action]
    Action $a=1$ (resp. $a=0$) is said to be \emph{motivated} by wishful thinking if $\mu_W(\rho) < \mu_B$ (resp. $\mu_W(\rho) > \mu_B$).
\end{definition}
That is, an action is motivated by wishful thinking if the set of beliefs supporting some action as optimal for a Bayesian agent is a strict subset of the set of (Bayesian) beliefs leading a wishful receiver to take the same action. Also, note that whenever a wishful receiver motivates an action $a$, such an agent takes this action with weakly greater probability than a Bayesian agent would under \emph{any} information structure.

Our next result shows that whether an action $a$ is motivated or not by a wishful receiver depends on two characteristics of his payoffs. We say that an action $a$ is the \textit{best-case action} if it yields the highest possible payoff in the receiver's decision problem, that is, if $\max_{\omega\in\Omega} u(a, \omega) \geq \max_{\omega\in\Omega} u(a', \omega)$, for $a \neq a'$. Furthermore, we say that an action $a$ is the \textit{riskier action} if it is the action with the highest payoff variability, measured in terms of absolute distance, in the receiver's decision problem, that is, if $\lvert u(a, 1) - u(a, 0)\rvert \geq \vert u(a', 1) - u(a', 0)\rvert$, for $a \neq a'$. Conversely, $a$ is the safer action if it is not the action with the highest payoff variability.


\begin{proposition}\label{thm:motivatedaction}
    Assume that the receiver wants to match his action with the state, that is $u(\omega,\omega)\geq u(1-\omega,\omega)$ for all $\omega\in \Omega$. Then, some action $a\in A$ is motivated by a wishful receiver if, and only if either $a$ is both the best-case action and the riskier action or:
    \begin{enumerate}[(i)]
        \item $a$ is the best-case action and the safer action and $\rho>\bar{\rho}$, or;
        \item $a$ is not the best-case action but is the riskier action and $\rho <\bar{\rho}$.
    \end{enumerate}
    where is the unique\footnote{We also prove the existence of $\bar{\rho}$ in \cref{secap:motivatedaction_proof}.} strictly positive wishfulness threshold $\bar{\rho}$ such that
    \begin{equation*}
        \mu_{W}(\bar{\rho})=\mu_{B}.
    \end{equation*}
\end{proposition}
\begin{proof}
    See \cref{secap:motivatedaction_proof}.
\end{proof}

Two key aspects of a wishful receiver's material payoff thus determine which action he motivates: \emph{the highest achievable payoff} as well as \emph{the payoff variability} for both actions. It is easy to grasp the importance of the highest payoff. Since the wishful thinker always distorts his beliefs in the direction of the most favorable outcome, in the limit, when there is no cost of distorting the Bayesian belief, Receiver would fully delude himself and always play the action that potentially yields such a payoff. The payoff variability, on the other hand, is precisely the wishful receiver's marginal psychological benefit from distorting his belief under action $a$. Hence, the higher the payoff variability associated with action $a$, the more the uncertainty about $\omega$ is relevant when such action is played and the bigger the marginal gain in anticipatory payoff the wishful thinker would get from distorting beliefs.

 \Cref{thm:motivatedaction} states that if the receiver wants to match the state---so that his decision problem is not trivial---then if an action $a$ is associated with both the highest payoff and the greatest payoff variability then it is always motivated. Moreover, if an action has either the highest payoff or the greatest payoff variability, then the degree of wishfulness of the receiver, $\rho$, determines which action he motivates. For sufficiently high degrees of wishfulness the action with the highest payoff is motivated, whereas for sufficiently low degrees of wishfulness the action with the greatest payoff variability that is motivated. The intuition is the following: for sufficiently high wishfulness, a wishful receiver can afford stronger overoptimism about the most desired outcome, thus motivating the action that potentially yields this outcome despite such action not being associated with the highest marginal psychological benefit. In contrast, for sufficiently low values of $\rho$, a wishful receiver cannot afford too much overoptimism about the most desired outcome. Hence, he prefers to distort beliefs at the margin that yields the highest marginal psychological benefit, such that the action associated with the highest payoff variability is motivated.

\subsection{Implications for the linear model}\label{sec:implications}
\Cref{thm:behavior_equivalence,thm:motivatedaction} establish how wishful receivers skew their behavior depending on their underlying preferences. We can now apply \cref{thm:motivatedaction} to our linear model.
\begin{corollary}\label{thm:implications_linear}
    In the linear model, for a receiver of type $(\gamma,\theta)$:
    \begin{enumerate}[(i)]
        \item If $\gamma = 0$, receiver regards $a=1$ as both the best-case and the riskier action. Such a receiver thus motivates $a=1$ for any $\rho > 0$.
        \item If $\gamma = 1$ and $\theta \in \mathopen[0,1/2\mathclose)$, receiver regards $a=1$ as both the best-case and the riskier action. Such a receiver thus motivates $a=1$ for any $\rho > 0$.
        \item If $\gamma = 1$ and $\theta =1/2$, receiver regards both actions as best-case and riskier. Such a receiver does not motivate any action.
        \item If $\gamma = 1$ and $\theta \in \mathopen(1/2,1\mathclose]$, receiver regards $a=0$ as both the best-case and the riskier action. Such a receiver thus motivates $a=0$ for any $\rho > 0$.
    \end{enumerate}
\end{corollary}
To grasp some intuition about \cref{thm:implications_linear} remark that:
\begin{equation*}
    u(a,\omega,0,\theta) = a(\omega-\theta)
\end{equation*}
while
\begin{equation*}
    u(a,\omega,1,\theta) = (1-\theta)a \omega + \theta (1-a)(1-\omega).
\end{equation*}
In the hypothetical world where the receiver is Bayesian, whether the value of the parameter $\gamma$---which simply determines the receiver's baseline payoff---would never change his optimal action, and could thus be normalized to zero \citep[as in][]{Kolotilin2017}. To see this, remark that two Bayesian receivers with a payoff function given by $u(a,\omega,0,\theta)$ or $u(a,\omega,1,\theta)$ would be indistinguishable from the sender's point of view, as the parameter $\theta$ would correspond in both cases to the belief indifference cutoff of the receiver. 

In contrast, when the receiver is wishful, normalizing the parameter $\gamma$ to zero is no longer without loss of generality as the uncertainty regarding the state is also relevant when the receiver takes action $a=0$. In particular, the exponential distortion of the payoff function induces the parameter $\gamma$ to determine which action is motivated by a wishful receiver as a function of $\theta$. In the specification where $\gamma=0$ all receiver types agree that the best possible outcome is to choose the action $a=1$ when the state is $\omega=1$ whatever the value of $\theta$. On the other hand, when $\gamma=1$, the best possible outcome for all receiver types such that $\theta$ is less than $1/2$ is to choose action $a=1$ when $\omega=1$, while the best possible outcome for all receiver types such that $\theta$ is greater than $1/2$ is to choose action $a=0$ when $\omega=0$.

As such, the probability $p$ that the parameter $\gamma$ is equal to $1$ captures the \emph{sender's uncertainty about the receiver's directional motive}, i.e., which actions he motivates. If $p=0$, the sender is sure that the receiver motivates action $a = 1$. As long as $p>0$, however, there exists a chance that the receiver motivates action $a=0$.

To illustrate \Cref{thm:implications_linear}, let us define the belief cutoff for a receiver of type $(\gamma, \theta)$ and wishfulness degree $\rho$ as follows:
\begin{equation}\label{eq:wishful_belief_cutoff_linear}
    \mu_W(\rho,\gamma,\theta) = \frac{e^{\gamma \rho \theta} - e^{(\gamma-1) \rho \theta}}{e^{\gamma \rho \theta} - e^{(\gamma-1) \rho \theta} + e^{\rho (1-\theta)} - 1},
\end{equation}
for every $(\gamma,\theta)\in T$, and $\rho\in\mathopen(0,+\infty\mathclose)$. The cutoff for a Bayesian receiver, on the other hand, is given by $\mu_B(\theta) = \theta$. \Cref{fig:cutoff} depicts how the wishful cutoff given by \cref{eq:wishful_belief_cutoff_linear} evolves as $\rho$ increases, for both $\gamma=0$ and $\gamma=1$.
\begin{figure}[h!]
    \centering
    \begin{subfigure}[t]{0.495\textwidth}
    \centering
    \begin{tikzpicture}[scale=0.9]
        \begin{axis}[
        axis x line=bottom,
        axis y line=left,
        xtick={0,1},
        xticklabels={$0$,$1$},
        ytick={1},
        yticklabels={$1$},
        every axis x label/.style={
        at={(ticklabel* cs:1.05)},
        anchor=west,
        },
        every axis y label/.style={
        at={(ticklabel* cs:1.05)},
        anchor=south,
        },
        domain=0:1,
        samples=100,
        xlabel={$\theta$},
        ylabel={$\mu_W(\rho,0,\cdot)$},
        ymin=0,
        ymax=1,
        legend cell align=left,
        legend pos=outer north east
        ]
        \addplot[smooth,thick,dotted] {1};
        \addplot[smooth,thick,dotted] coordinates{(1,0)(1,1)};
        \addplot[smooth,dotted] coordinates{(0,1/2)(1/2,1/2)};
        \addplot[smooth,dotted] coordinates{(1/2,0)(1/2,1/2)};
        \addplot[smooth,mBlue!20!white] {x};
        \addplot[smooth,thick,mBlue!25!white] {(exp(2*x)-1)/(exp(2)-1)};
        \addplot[smooth,thick,mBlue!50!white] {(exp(4*x)-1)/(exp(4)-1)};
        \addplot[smooth,thick,mBlue!75!white] {(exp(7*x)-1)/(exp(7)-1)};
        \addplot[smooth,thick,mBlue!100!white] {(exp(10*x)-1)/(exp(10)-1)};
        \end{axis}
    \end{tikzpicture}
    \caption{Indifference belief cutoff when $\gamma=0$}
\end{subfigure}
\begin{subfigure}[t]{0.495\textwidth}
    \centering
    \begin{tikzpicture}[scale=0.9]
        \begin{axis}[
        axis x line=bottom,
        axis y line=left,
        xtick={0,1/2,1},
        xticklabels={$0$,$\frac{1}{2}$,$1$},
        ytick={1/2,1},
        yticklabels={$\frac{1}{2}$,$1$},
        every axis x label/.style={
        at={(ticklabel* cs:1.05)},
        anchor=west,
        },
        every axis y label/.style={
        at={(ticklabel* cs:1.05)},
        anchor=south,
        },
        domain=0:1,
        samples=100,
        xlabel={$\theta$},
        ylabel={$\mu_W(\rho,0,\cdot)$},
        ymin=0,
        ymax=1,
        legend cell align=left,
        legend pos=outer north east
        ]
        \addplot[smooth,thick,dotted] {1};
        \addplot[smooth,thick,dotted] coordinates{(1,0)(1,1)};
        \addplot[smooth,dotted] coordinates{(0,1/2)(1/2,1/2)};
        \addplot[smooth,dotted] coordinates{(1/2,0)(1/2,1/2)};
        \addplot[smooth,mBlue!20!white] {x};
        \addplot[smooth,thick,mBlue!25!white] {(exp(2*x)-1)/(exp(2*x)+exp(2*(1-x))-2)};
        \addplot[smooth,thick,mBlue!50!white] {(exp(4*x)-1)/(exp(4*x)+exp(4*(1-x))-2)};
        \addplot[smooth,thick,mBlue!75!white] {(exp(7*x)-1)/(exp(7*x)+exp(7*(1-x))-2)};
        \addplot[smooth,thick,mBlue!100!white] {(exp(10*x)-1)/(exp(10*x)+exp(10*(1-x))-2)};
        \end{axis}
    \end{tikzpicture}
    \caption{Indifference belief cutoff when $\gamma=1$.}
\end{subfigure}
\caption{Belief indifference cutoff as a function of receiver's type $(\gamma,\theta)$ as $\rho$ increases. Darker lines correspond to higher values of $\rho$.}
\label{fig:cutoff}
\end{figure}
First of all, we get back to the Bayesian case when the cost of belief distortion becomes arbitrarily large, since $\lim_{\rho\to 0^{+}} \mu_{W}(\rho,\gamma,\theta)=\theta=\mu_{B}(\theta)$ for any $(\gamma,\theta)\in T$.\footnote{All the proofs regarding the properties of the wishful indifference cutoff $\mu_{W}(\rho,\gamma,\theta)$ defined in \cref{eq:wishful_belief_cutoff_linear} can be found in \cref{secap:proof_indif_cutoff}.} In the case where $\gamma=0$, the function $\mu_W(\rho,0,\cdot)$ is strictly convex and is such that $\mu_W(\rho,0,0)=0$ and $\mu_W(\rho,0,1)=1$. Therefore, $\mu_W(\rho,0,\theta)<\theta=\mu_{B}(\theta)$ for all $\theta\in\mathopen[0,1\mathclose]$, which implies that all receiver types such that $\gamma=0$ motivate action $a=1$. Conversely, when $\gamma=1$, the function $\mu_W(\rho,0,\cdot)$ has a unique inflection point at $\theta=1/2$, is strictly convex on the interval $\mathopen[0,1/2\mathclose)$, is strictly concave on the interval $\mathopen(1/2,1\mathclose]$ and is such that $\mu_W(\rho,0,0)=0$, $\mu_W(0,1/2,\rho)=1/2$ and $\mu_W(\rho,0,1)=1$ for any $\rho\in\mathopen(0,+\infty\mathclose)$. Therefore, $\mu_W(\rho,0,\theta)<\theta=\mu_{B}(\theta)$ for all $\theta\in\mathopen[0,1/2\mathclose)$ while $\mu_W(\rho,0,\theta)>\theta=\mu_{B}(\theta)$ for all $\theta\in\mathopen(1/2,1\mathclose]$, which implies that receiver types such that $\gamma=1$ and $\theta\in\mathopen[0,1/2\mathclose)$ motivate action $a=1$ while receiver types such that $\gamma=1$ and $\theta\in\mathopen(1/2,1\mathclose]$ motivate action $a=0$.

\subsection{Restating the sender's problem}

Importantly, \cref{thm:implications_linear} allows us to state the \emph{non-Bayesian linear} persuasion problem of the sender into a \emph{non-linear Bayesian} persuasion problem. To do so, let us state a useful intermediary result.
\begin{lemma}\label{thm:inv_bij}
    The function $\theta\mapsto \mu_{W}(\rho,\gamma,\theta)$ admits an inverse bijection $\mu\mapsto \vartheta(\mu,\rho,\gamma)$ such that:
    \begin{enumerate}[(i)]
        \item If $\gamma=0$ then
        \begin{equation*}
            \vartheta(\mu,\rho,0)=\frac{1}{\rho}\ln\bigl(1+(e^{\rho}-1)\mu\bigr),
        \end{equation*}
        for all $\mu\in\mathopen[0,1\mathclose]$ and $\rho\in\mathopen(0,+\infty\mathclose)$.
        \item If $\gamma=0$ then
        \begin{equation*}
            \vartheta(\mu,\rho,1)=\frac{1}{\rho}\ln\Biggl(\frac{1-2\mu+\sqrt{1+4(e^{\rho}-1)\mu(1-\mu)}}{2(1-\mu)}\Biggr),
        \end{equation*}
        for all $\mu\in\mathopen[0,1\mathclose]$ and $\rho\in\mathopen(0,+\infty\mathclose)$.
    \end{enumerate}
\end{lemma}
The proof is straightforward and omitted. Essentially, the function $\vartheta(\cdot,\rho,\gamma)$ is symmetric to $\mu_{W}(\rho,\gamma,\cdot)$ along the $45^{\circ}$ line. Therefore it is always strictly increasing and, when $\gamma=0$, it is strictly concave and such that $\vartheta(0,\rho,0)=0$ and $\vartheta(1,\rho,0)=1$, while when $\gamma=1$, it is strictly concave-convex with a unique inflection point at $\theta=1/2$ and such that $\vartheta(0,\rho,1)=0$, $\vartheta(1/2,\rho,1)=1/2$ and $\vartheta(1,\rho,1)=1$. An example is depicted on \cref{fig:inv_bij}.
\begin{figure}[h!]
    \centering
    \begin{subfigure}[t]{0.495\textwidth}
    \centering
    \begin{tikzpicture}[scale=0.9]
        \begin{axis}[
        axis x line=bottom,
        axis y line=left,
        xtick={0,1},
        xticklabels={$0$,$1$},
        ytick={1},
        yticklabels={$1$},
        every axis x label/.style={
        at={(ticklabel* cs:1.05)},
        anchor=west,
        },
        every axis y label/.style={
        at={(ticklabel* cs:1.05)},
        anchor=south,
        },
        domain=0:1,
        samples=100,
        xlabel={$\mu$},
        ylabel={$\vartheta(\cdot,\rho,0)$},
        ymin=0,
        ymax=1,
        legend cell align=left,
        legend pos=outer north east
        ]
        \addplot[smooth,thick,dotted] {1};
        \addplot[smooth,thick,dotted] coordinates{(1,0)(1,1)};
        \addplot[smooth,mBlue!20!white] {x};
        \addplot[smooth,thick,mBlue!25!white] {(1/2)*ln(1+(e^2-1)*x};
        \addplot[smooth,thick,mBlue!50!white] {(1/4)*ln(1+(e^4-1)*x};
        \addplot[smooth,thick,mBlue!75!white] {(1/7)*ln(1+(e^7-1)*x};
        \addplot[smooth,thick,mBlue!100!white] {(1/10)*ln(1+(e^10-1)*x};
        \end{axis}
    \end{tikzpicture}
    \caption{Indifference type cutoff when $\gamma=0$.}
\end{subfigure}
\begin{subfigure}[t]{0.495\textwidth}
    \centering
    \begin{tikzpicture}[scale=0.9]
        \begin{axis}[
        axis x line=bottom,
        axis y line=left,
        xtick={0,1},
        xticklabels={$0$,$1$},
        ytick={1},
        yticklabels={$1$},
        every axis x label/.style={
        at={(ticklabel* cs:1.05)},
        anchor=west,
        },
        every axis y label/.style={
        at={(ticklabel* cs:1.05)},
        anchor=south,
        },
        domain=0:1,
        samples=100,
        xlabel={$\mu$},
        ylabel={$\vartheta(\cdot,\rho,1)$},
        ymin=0,
        ymax=1,
        legend cell align=left,
        legend pos=outer north east
        ]
        \addplot[smooth,thick,dotted] {1};
        \addplot[smooth,thick,dotted] coordinates{(1,0)(1,1)};
        \addplot[smooth,dotted] coordinates{(0,1/2)(1/2,1/2)};
        \addplot[smooth,dotted] coordinates{(1/2,0)(1/2,1/2)};
        \addplot[smooth,mBlue!20!white] {x};
        \addplot[smooth,thick,mBlue!25!white] {(1/2)*ln((1-2*x+sqrt(1+4*(e^2-1)*x*(1-x)))/(2*(1-x)))};
        \addplot[smooth,thick,mBlue!50!white] {(1/4)*ln((1-2*x+sqrt(1+4*(e^4-1)*x*(1-x)))/(2*(1-x)))};
        \addplot[smooth,thick,mBlue!75!white] {(1/7)*ln((1-2*x+sqrt(1+4*(e^7-1)*x*(1-x)))/(2*(1-x)))};
        \addplot[smooth,thick,mBlue!100!white] {(1/10)*ln((1-2*x+sqrt(1+4*(e^10-1)*x*(1-x)))/(2*(1-x)))};
        \end{axis}
    \end{tikzpicture}
    \caption{Indifference type cutoff when $\gamma=1$.}
\end{subfigure}
\caption{Indifferent receiver type as a function of $\mu$ as $\rho$ increases. Darker lines correspond to higher values of $\rho$.}
\label{fig:inv_bij}
\end{figure}
Intuitively, for any level of wishfulness $\rho$, the function $\vartheta(\cdot,\rho,\gamma)$ pins-down the receiver types who are indifferent between the two actions when the sender's information policy induces the Bayesian posterior $\mu$.

Conveniently, \cref{thm:inv_bij} implies that the optimal action of a wishful thinker is given by
\begin{equation*}
    a(\mu,\rho,\gamma,\theta)=\mathds{1}_{\theta\leq\vartheta(\mu,\rho,\gamma)},
\end{equation*}
for every $\mu\in\mathopen[0,1\mathclose]$, $\rho\in\mathopen(0,+\infty\mathclose)$ and $(\gamma,\theta)\in T$, so that the expected action of the receiver conditional on $\gamma$ is given by
\begin{align*}
    \int_{0}^{1} \mathds{1}_{\mu\geq \mu_{W}(\rho,\theta,\gamma)}\, f(\theta) \, \mathrm{d}\theta &=\int_{0}^{1} \mathds{1}_{\theta\leq \vartheta(\mu,\rho,\gamma)}\, f(\theta) \, \mathrm{d}\theta \\
     &=\int_{0}^{\vartheta(\mu,\rho,\gamma)} f(\theta) \, \mathrm{d}\theta \\
    &=F\big(\vartheta(\mu,\rho,\gamma)\big).
\end{align*}
for any $\mu\in\mathopen[0,1\mathclose]$, $\rho\in\mathopen(0,+\infty\mathclose)$ and $\gamma\in \{0,1\}$. This implies that the indirect utility function of the sender at any posterior belief $\mu$ is equal to the probability that the receiver's type is such that $\theta$ lies below the indifference cutoff $\vartheta(\mu,\rho,\gamma)$, given by
\begin{equation*}
    V(\mu,\rho)= p F\big(\vartheta(\mu,\rho,1)\big) + (1-p)  F\big(\vartheta(\mu,\rho,0)\big),
\end{equation*}
for every $\mu\in\mathopen[0,1\mathclose]$ and $\rho\in\mathopen(0,+\infty\mathclose)$. The sender's information design problem thus consists in solving the following optimization program:
\begin{equation*}
    \max_{\pi\in\Pi(\mu_{0})} \; \int_{\mathopen[0,1\mathclose]} \Bigl(p F\big(\vartheta(\mu,\rho,1)\big) + (1-p)  F\big(\vartheta(\mu,\rho,0)\big)\Bigr) \, \pi(\mathrm{d}\mu).
\end{equation*}

\subsection{Main result}

Given the results established in the previous sections, we can move on to establishing sender's optimal information structure. In order to do that, we categorize information structures following \cite{Kolotilin2022}.

An information structure $(\sigma, S)$ exhibits \textit{upper-censorship} if it generates one signal $s=0$ that fully discloses state $\omega=0$ and one signal $s=1$ that pools together states $\omega=0$ and $\omega=1$.\footnote{As the receiver in our model faces a binary decision problem, it is without loss of generality to focus on information structures for which the signal space $S$ only includes two elements.} Conversely, an information structure $(\sigma, S)$ exhibits \textit{lower-censorship} if it generates one signal $s=0$ that pools together states $\omega=0$ and $\omega=1$ and one signal $s=1$ that fully discloses state $\omega=1$.

Importantly, remark that in a setting with a Bayesian receiver but otherwise identical to ours, an \emph{upper-censorship} policy would be uniquely optimal. Indeed, since the optimal action of a Bayesian receiver is given by $a(\mu)=\mathds{1}_{\mu\geq \theta}$ irrespective of whether $\gamma=0$ or $\gamma=1$, the sender's indirect utility function in that case is given by
\begin{align*}
    V(\mu)&=\int_{0}^{1} \mathds{1}_{\theta\leq \mu} \, f(\theta) \, \mathrm{d}\theta\\
    &=\int_{0}^{\mu} f(\theta) \, \mathrm{d}\theta\\
    &=F(\mu),
\end{align*}
for all $\mu\in\mathopen[0,1\mathclose]$. We illustrate on \Cref{fig:Bayesian_optimum} the Bayesian-optimal information policy together with the sender's optimal payoff.
\begin{figure}[h!]
    \centering
    \begin{tikzpicture}[scale=0.9]
        \begin{axis}[
        axis x line=bottom,
        axis y line=left,
        xtick={0,0.35,0.676,1},
        xticklabels={$0$,$\mathcolor{mOrange}{\mu_{0}}$,$\mathcolor{mOrange}{\mu^{*}}$,$1$},
        ytick={1},
        yticklabels={$1$},
        every axis x label/.style={
        at={(ticklabel* cs:1.05)},
        anchor=west,
        },
        every axis y label/.style={
        at={(ticklabel* cs:1.05)},
        anchor=south,
        },
        domain=0:1,
        samples=100,
        xlabel={$\mu$},
        ylabel=\empty,
        ymin=0,
        ymax=1.01,
        legend cell align=left,
        legend pos=outer north east
        ]
        \addplot[smooth,thick,dotted] {1};
        \addplot[smooth,thick,dotted] coordinates{(1,0)(1,1)};
        
        \addplot[smooth,very thick,mBlue]{(1/(1+exp(-(x-0.5)/0.1))-1/(1+exp(-(0-0.5)/0.1)))/(1/(1+exp(-(1-0.5)/0.1))-1/(1+exp(-(0-0.5)/0.1)))};
        
        \addplot[smooth,very thick,mOrange,domain=0.676:1]{(1/(1+exp(-(x-0.5)/0.1))-1/(1+exp(-(0-0.5)/0.1)))/(1/(1+exp(-(1-0.5)/0.1))-1/(1+exp(-(0-0.5)/0.1)))};
        \addplot[smooth,very thick,mOrange,domain=0:0.676]{1.26942145656*x};
        \node (source_0) at (axis cs:0.36,0){};
        \node (source_1) at (axis cs:0.34,0){};
        \node (destination_0) at (axis cs:-0.01,0){};
        \node (destination_1) at (axis cs:0.68,0){};
        \draw[-stealth, thick,mOrange,opacity=0.50](source_0) to[bend right] (destination_0);
        \draw[-stealth, thick, mOrange,opacity=0.50](source_1) to[bend left] (destination_1);
        \node at (axis cs:0.35,0.75){$\mathcolor{mOrange}{\mathrm{cav}(V)(\mu)}$};
        \node at (axis cs:0.5,0.25){$\mathcolor{mBlue}{V(\mu)}$};
        \addplot[smooth,dotted,thick,mOrange] coordinates{(0.676,0)(0.676,0.858002)};
        \end{axis}
    \end{tikzpicture}
    \caption{Bayesian-optimal information policy.}
    \label{fig:Bayesian_optimum}
\end{figure}
Since the density $f$ is log-concave, the cumulative distribution function $F$ must be convex-concave on the interval $\mathopen[0,1\mathclose]$. Therefore, Corollary 2 in \cite{Kamenica2011} implies that the sender's Bayesian-optimal information policy is supported on $\{0,\mu^{*}\}$ whenever $\mu_{0}\in\mathopen[0,\mu^{*})$ and is supported on $\{\mu_{0}\}$ whenever $\mu_{0}\in\mathopen[\mu^{*},1\mathclose]$. This implies that the corresponding optimal information structure is an upper-censorship generating one signal $s=0$ that fully discloses state $\omega=0$ and one signal $s=1$ that pools together states $\omega=0$ and $\omega=1$ so as to induce the receiver to hold the posterior belief $\mu^{*}$.

In contrast, our main result establishes that optimally persuading a wishful receiver might require using a \emph{lower-censorship} policy even if $f$ is log-concave.
\begin{theorem}[Optimal information policy]\label{thm:opt_info}
    If sender faces no uncertainty about which action the receiver motivates, i.e., if $p=0$, then there exists a unique wishfulness cutoff $\Tilde{\rho}$ such that if $\rho<\Tilde{\rho}$ then the sender's optimal policy is an \emph{upper-censorship policy}, and if $\rho\geq \Tilde{\rho}$ the sender's optimal policy is \emph{non-informative}. Conversely, if the sender is uncertain about which action the receiver motivates, i.e., if $p>0$, then there exists two wishfulness cutoffs $\ubar{\rho}$ and $\bar{\rho}$ such that $0<\ubar{\rho}<\bar{\rho}$ and that:
    \begin{enumerate}[(i)]
        \item If $\rho\leq\ubar{\rho}$ the optimal information policy of the sender is an \emph{upper-censorship policy}.
        \item If $\rho\geq\ubar{\rho}$ the optimal information policy of the sender is a \emph{lower-censorship policy}.
    \end{enumerate}
\end{theorem}
\begin{proof}
    See \Cref{secap:proof_opt_info}.
\end{proof}


In order to understand the intuition behind \Cref{thm:opt_info}, let's first note that sender's information policy should generate two possible signal realizations, each of which leads to some posterior belief $\mu$. Each posterior realization $\mu$ is associated with a given probability of successfully persuading the receiver. Receiver indeed chooses the sender's desired action if $\mu \geq \mu_W(\gamma, \theta, \rho)$, which happens with probability $V(\mu, \rho)$. Let us call ``good news'' the signal suggestive of $\omega = 1$ (which leads to a higher $\mu$ and is associated with a higher probability of inducing $a=1$), and ``bad news'' the other one. Bayes-plausibility ties down the posteriors associated with both signals with the probability of each signal realization, such that making some signal more indicative of a given state decreases the (ex-ante) probability of such signal being realized.
\begin{figure}[h!]
    \centering
    \begin{tikzpicture}[scale=0.9]
        \begin{axis}[
        axis x line=bottom,
        axis y line=left,
        xtick={0,0.17,0.355,1},
        xticklabels={$0$,$\mathcolor{mOrange}{\mu_{0}}$,$\mathcolor{mOrange}{\mu^{*}}$,$1$},
        ytick={1},
        yticklabels={$1$},
        every axis x label/.style={
        at={(ticklabel* cs:1.05)},
        anchor=west,
        },
        every axis y label/.style={
        at={(ticklabel* cs:1.05)},
        anchor=south,
        },
        domain=0:1,
        samples=100,
        xlabel={$\mu$},
        ylabel=\empty,
        ymin=0,
        ymax=1.01,
        legend cell align=left,
        legend pos=outer north east
        ]
        \addplot[smooth,thick,dotted] {1};
        \addplot[smooth,thick,dotted] coordinates{(1,0)(1,1)};
        
        \addplot[smooth,very thick,dashed,mBlue!33!white]{(1/(1+exp(-(x-0.5)/0.1))-1/(1+exp(-(0-0.5)/0.1)))/(1/(1+exp(-(1-0.5)/0.1))-1/(1+exp(-(0-0.5)/0.1)))};
        
        \addplot[smooth,very thick,dashed,mOrange!33!white,domain=0.676:1]{(1/(1+exp(-(x-0.5)/0.1))-1/(1+exp(-(0-0.5)/0.1)))/(1/(1+exp(-(1-0.5)/0.1))-1/(1+exp(-(0-0.5)/0.1)))};
        \addplot[smooth,very thick,dashed,mOrange!33!white,domain=0:0.676]{1.26942145656*x};
        
        
        \addplot[smooth,very thick,mBlue!100!white]{(1/(1+exp(-((1/2)*ln(1+(e^2-1)*x)-0.5)/0.1))-1/(1+exp(-(0-0.5)/0.1)))/(1/(1+exp(-(1-0.5)/0.1))-1/(1+exp(-(0-0.5)/0.1)))};

        
        \addplot[smooth,very thick,mOrange!100!white,domain=0.355:1]{(1/(1+exp(-((1/2)*ln(1+(e^2-1)*x)-0.5)/0.1))-1/(1+exp(-(0-0.5)/0.1)))/(1/(1+exp(-(1-0.5)/0.1))-1/(1+exp(-(0-0.5)/0.1)))};
        \addplot[smooth,very thick,mOrange!100!white,domain=0:0.355]{2.0177852754*x};
        \node (source_0) at (axis cs:0.17,0){};
        \node (source_1) at (axis cs:0.17,0){};
        \node (destination_0) at (axis cs:-0.01,0){};
        \node (destination_1) at (axis cs:0.37,0){};
        \draw[-stealth, thick,mOrange,opacity=0.50](source_0) to[bend right] (destination_0);
        \draw[-stealth, thick, mOrange,opacity=0.50](source_1) to[bend left] (destination_1);
        \node[left] at (axis cs:0.35,0.75){$\mathcolor{mOrange}{\mathrm{cav}(V)(\mu)}$};
        \node at (axis cs:0.28,0.20){$\mathcolor{mBlue}{V(\mu)}$};
        \addplot[smooth,dotted,thick,mOrange] coordinates{(0.355,0)(0.355,0.71818)};
        \end{axis}
    \end{tikzpicture}
    \caption{Wishful-optimal information policy when $\gamma=0$ and $\rho>0$ in comparison to the Bayesian-optimal one.}
    \label{fig:gamma0}
\end{figure}

Upper-censorship policies' appeal stems from the fact that making bad news conclusive of state $\omega = 0$ makes the realization of good news ex-ante more likely. One can then adjust how suggestive of $\omega=1$ good news is: making it more suggestive increases the probability that such signal will induce the desired action, but also decreases the ex-ante likelihood of good news being realized. An optimal upper-censorship policy fine-tunes the meaning of good news by balancing such trade-off.

For low wishfulness degree $\rho$, each receiver type's cutoff $\mu_W(\gamma, \theta, \rho)$ is close to $\theta$. The log-concavity of the distribution of $\theta$ then implies that upper censorship policies are optimal since, by making bad news conclusive of $\omega=0$, the gain in ex-ante likelihood of good news more than compensates for the loss in the probability of inducing the preferred action under bad news. 

But what about higher levels of $\rho$? If every receiver type motivates sender's preferred action, then \textit{every} type's cutoff $\mu_W$ decreases as $\rho$ increases. Furthermore, this shift is such that the underlying distribution of cutoffs $\mu_W$ remains log-concave, implying that it is still advantageous to have conclusive bad news. The shift in indifference cutoffs also means that every posterior realization is now associated with a higher probability of inducing sender's preferred action. In other terms, $V(\mu,\rho')$ strictly dominates $V(\mu,\rho)$ in terms of first-order stochastic dominance when $\rho'>\rho$. That implies that the trade-off implied in the fine tuning of good news can be balanced by providing less suggestive evidence of $\omega=1$. This case is pictured in \cref{fig:gamma0}.

The second part of point (i) in \cref{thm:opt_info} indicates that if the receiver's degree of wishfulness becomes sufficiently high then indifference cutoffs become so concentrated at the bottom end of the interval $\mathopen[0,1\mathclose]$ that all receiver types would still be convinced to opt for the sender's preferred action under no information disclosure, i.e., $V(\mu,\rho)$ would be concave in $\mu$. Although $a=1$ is the motivated action when $p = 0$, it is not difficult to show that if the sender was instead certain that the receiver motivated action was $a=0$ then the optimal information structure would still exhibit upper-censorship but would be more informative than the Bayesian-optimal one as there would be a strictly decreasing first-order stochastic dominance shift between $V(\mu,\rho)$ and $V(\mu,\rho')$ as $\rho$ shifts to $\rho'>\rho$.

\begin{figure}[h!]
    \centering
    \begin{subfigure}[t]{0.495\textwidth}
        \centering
        \begin{tikzpicture}[scale=0.9]
            \begin{axis}[
            axis x line=bottom,
            axis y line=left,
            xtick={0,0.35,0.7333,1},
            xticklabels={$0$,$\mathcolor{mOrange}{\mu_{0}}$,$\mathcolor{mOrange}{\mu^{*}}$,$1$},
            ytick={1},
            yticklabels={$1$},
            every axis x label/.style={
            at={(ticklabel* cs:1.05)},
            anchor=west,
            },
            every axis y label/.style={
            at={(ticklabel* cs:1.05)},
            anchor=south,
            },
            domain=0:1,
            samples=100,
            xlabel={$\mu$},
            ylabel=\empty,
            ymin=0,
            ymax=1.01,
            legend cell align=left,
            legend pos=outer north east
            ]
            \addplot[smooth,thick,dotted] {1};
            \addplot[smooth,thick,dotted] coordinates{(1,0)(1,1)};
            
            \addplot[smooth,very thick,dashed,mBlue!33!white]{(1/(1+exp(-(x-0.5)/0.1))-1/(1+exp(-(0-0.5)/0.1)))/(1/(1+exp(-(1-0.5)/0.1))-1/(1+exp(-(0-0.5)/0.1)))};
            
            \addplot[smooth,very thick,dashed,mOrange!33!white,domain=0.676:1]{(1/(1+exp(-(x-0.5)/0.1))-1/(1+exp(-(0-0.5)/0.1)))/(1/(1+exp(-(1-0.5)/0.1))-1/(1+exp(-(0-0.5)/0.1)))};
            \addplot[smooth,very thick,dashed,mOrange!33!white,domain=0:0.676]{1.26942145656*x};

            \addplot[smooth,very thick,mBlue!100!white]{(1/(1+e^(-(((1/2.5)*ln((1-2*x+sqrt(1+4*(e^(2.5)-1)*x*(1-x)))/(2*(1-x))))-0.5)/0.1))-1/(1+e^(-(0-0.5)/0.1)))/(1/(1+e^(-(1-0.5)/0.1))-1/(1+e^(-(0-0.5)/0.1)))};


            \addplot[smooth,very thick,mOrange!100!white,domain=0.7333:1]{(1/(1+e^(-(((1/2.5)*ln((1-2*x+sqrt(1+4*(e^(2.5)-1)*x*(1-x)))/(2*(1-x))))-0.5)/0.1))-1/(1+e^(-(0-0.5)/0.1)))/(1/(1+e^(-(1-0.5)/0.1))-1/(1+e^(-(0-0.5)/0.1)))};
            \addplot[smooth,very thick,mOrange!100!white,domain=0:0.7333]{1.10410756535*x};

            \node (source_0) at (axis cs:0.35,0){};
            \node (source_1) at (axis cs:0.35,0){};
            \node (destination_0) at (axis cs:-0.01,0){};
            \node (destination_1) at (axis cs:0.74,0){};
            \draw[-stealth, thick,mOrange,opacity=0.50](source_0) to[bend right] (destination_0);
            \draw[-stealth, thick, mOrange,opacity=0.50](source_1) to[bend left] (destination_1);
            \node[left] at (axis cs:0.47,0.65){$\mathcolor{mOrange}{\mathrm{cav}(V)(\mu)}$};
            \node at (axis cs:0.53,0.3){$\mathcolor{mBlue}{V(\mu)}$};
            \addplot[smooth,dotted,thick,mOrange] coordinates{(0.7333,0)(0.7333,0.80957)};
            
            \end{axis}
        \end{tikzpicture}
        \caption{Optimal information policy when $\rho<\ubar{\rho}$.}
        \label{fig:upper_censorship}
    \end{subfigure}
    \begin{subfigure}[t]{0.495\textwidth}
        \centering
        \begin{tikzpicture}[scale=0.9]
            \begin{axis}[
            axis x line=bottom,
            axis y line=left,
            xtick={0,0.1572,0.55,1},
            xticklabels={$0$,$\mathcolor{mOrange}{\hat{\mu}}$,$\mathcolor{mOrange}{\mu_{0}}$,$1$},
            ytick={1},
            yticklabels={$1$},
            every axis x label/.style={
            at={(ticklabel* cs:1.05)},
            anchor=west,
            },
            every axis y label/.style={
            at={(ticklabel* cs:1.05)},
            anchor=south,
            },
            domain=0:1,
            samples=100,
            xlabel={$\mu$},
            ylabel=\empty,
            ymin=0,
            ymax=1.01,
            legend cell align=left,
            legend pos=outer north east
            ]
            \addplot[smooth,thick,dotted] {1};
            \addplot[smooth,thick,dotted] coordinates{(1,0)(1,1)};
            
            \addplot[smooth,very thick,dashed,mBlue!33!white]{(1/(1+exp(-(x-0.5)/0.1))-1/(1+exp(-(0-0.5)/0.1)))/(1/(1+exp(-(1-0.5)/0.1))-1/(1+exp(-(0-0.5)/0.1)))};
            
            \addplot[smooth,very thick,dashed,mOrange!33!white,domain=0.676:1]{(1/(1+exp(-(x-0.5)/0.1))-1/(1+exp(-(0-0.5)/0.1)))/(1/(1+exp(-(1-0.5)/0.1))-1/(1+exp(-(0-0.5)/0.1)))};
            \addplot[smooth,very thick,dashed,mOrange!33!white,domain=0:0.676]{1.26942145656*x};

            \addplot[smooth,very thick,mBlue!100!white]{(1/(1+e^(-(((1/9)*ln((1-2*x+sqrt(1+4*(e^(9)-1)*x*(1-x)))/(2*(1-x))))-0.5)/0.1))-1/(1+e^(-(0-0.5)/0.1)))/(1/(1+e^(-(1-0.5)/0.1))-1/(1+e^(-(0-0.5)/0.1)))};

            \addplot[smooth,very thick,mOrange!100!white,domain=0:0.1572]{(1/(1+e^(-(((1/9)*ln((1-2*x+sqrt(1+4*(e^(9)-1)*x*(1-x)))/(2*(1-x))))-0.5)/0.1))-1/(1+e^(-(0-0.5)/0.1)))/(1/(1+e^(-(1-0.5)/0.1))-1/(1+e^(-(0-0.5)/0.1)))};
            \addplot[smooth,very thick,mOrange!100!white,domain=0.1572:1]{0.8523*(x-0.1572)+0.2818};

            \node (source_0) at (axis cs:0.55,0){};
            \node (source_1) at (axis cs:0.55,0){};
            \node (destination_0) at (axis cs:1.01,0){};
            \node (destination_1) at (axis cs:0.14,0){};
            \draw[-stealth, thick,mOrange,opacity=0.50](source_0) to[bend left] (destination_0);
            \draw[-stealth, thick, mOrange,opacity=0.50](source_1) to[bend right] (destination_1);
            \node[left] at (axis cs:0.47,0.65){$\mathcolor{mOrange}{\mathrm{cav}(V)(\mu)}$};
            \node at (axis cs:0.53,0.3){$\mathcolor{mBlue}{V(\mu)}$};
            \addplot[smooth,dotted,thick,mOrange] coordinates{(0.1572,0)(0.1572,0.2817)};
            
            \end{axis}
        \end{tikzpicture}
        \caption{Optimal information policy when $\rho>\bar{\rho}$.}
        \label{fig:lower_censorship}
    \end{subfigure}
    \caption{Wishful-optimal information policy when $\gamma=1$ and $\rho>0$, in comparison to the Bayesian-optimal one}
    \label{fig:gamma1}
\end{figure}

Things, however, become different if receiver types motivate different actions. The cutoff of receiver types that motivate $a=1$ decreases, meaning that they require less evidence of $\omega=1$ in order to take sender's preferred action, whereas the cutoff of types that motivate $a=0$ increases, meaning that more evidence of $\omega=1$ is needed for them in order to take the sender's preferred action.

Intuitively, this polarization has two consequences for the sender.  The first one is that it is no longer advantageous to make bad news conclusive of $\omega=0$. This is so because, as $\rho$ becomes large enough, highly-suggestive (but non-conclusive) bad news would be enough to induce receiver types that motivate action $a=1$ to take sender's preferred action. As such, the loss from not persuading those receiver types under bad news would be greater than the gain in ex-ante probability of good news that could be obtained if bad news were conclusive. The second implication is that, as $\rho$ increases, good news need to be more and more suggestive in order to persuade receiver types that motivate action $a=0$. This leads good news to be conclusive under the optimal information policy. Therefore, the optimal information policy when there is uncertainty about receiver's motives is one that induces sender's preferred action with positive probability after \emph{any} signal realization.

Formally, the \emph{polarization} in the indifference cutoffs between receiver types for which $\theta<1/2$ and $\theta>1/2$ causes a \emph{dispersion} shift in the function $V(\mu,\rho)$ as $\rho$ increases.\footnote{This type of dispersion shift in the indirect utility function $V(\mu,\rho)$ corresponds to a \emph{rotation} shift in the sense of \cite{Johnson2006}.} When $\rho<\bar{\rho}$, this increased dispersion translates into $V(\mu,\rho)$ rotating around the point $(1/2,V(1/2,\rho))$ while staying S-shaped, so the optimal information policy still exhibits upper-censorship, as illustrated on \cref{fig:upper_censorship}.\footnote{Equivalently, since $V(\cdot,\rho)$ is a cumulative distribution function for any $\rho$, its density function $v(\mu,\rho)=\frac{\partial V}{\partial\mu}(\mu,\rho)$ stays \emph{single-peaked} but has fatter tails.} When $\rho>\bar{\rho}$, conversely, the dispersion in cutoffs becomes so high that $V(\mu,\rho)$ rotates around the point $(1/2,V(1/2,\rho))$ while becoming inverse-S-shaped.\footnote{Equivalently, its density function $v(\mu,\rho)=\frac{\partial V}{\partial\mu}(\mu,\rho)$ becomes \emph{single-dipped}.} This implies that the optimal information policy is now supported on $\{\mu_{0}\}$ if $\mu_{0}<\hat{\mu}$ and on $\{\hat{\mu},1\}$ if $\mu_{0}>\hat{\mu}$, as illustrated on \cref{fig:lower_censorship}.

That information policy corresponds to a lower-censorship information structure generating one signal $s=0$ that pools together states $\omega=0$ and $\omega=1$ so as to induce the receiver to hold the posterior belief $\hat{\mu}$ and one signal $s=1$ that fully discloses state $\omega=1$.
 
It is worth noting that \cref{thm:opt_info} does not rely on the particularities of our payoff specification. For instance, the fact that the best-case action always coincides with the riskier action is inconsequential for the reversal pointed out by our result: since the best-case action is always motivated for high enough values of $\rho$, the lower-censorship structure emerges as optimal for sufficiently high $\rho$ whenever different types disagree on the best-case action, regardless of what they perceive as the riskier action.

\bibliographystyle{apalike}
\bibliography{biblio_merged}

\clearpage

\appendix
\begin{center}
    \Huge \textbf{Mathematical Appendix}
\end{center}

\small

\section{Proof for \cref{thm:behavior_equivalence}}\label{secap:behavior_equivalence_proof}

Let $\Omega$ be a Polish space and let $\Delta(\Omega)$ be the set of Borel probability measures on $\Omega$. For any probability measures $\eta$ and $\mu$ belonging to $\Delta(\Omega)$ and such that $\eta$ is absolutely continuous with respect to $\mu$, the Kullback-Leibler divergence of $\eta$ with respect to $\mu$ is defined by
\begin{equation*}
    C(\eta,\mu)=\int_{\Omega} \ln\left(\frac{\mathrm{d}\eta}{\mathrm{d}\mu}(\omega)\right) \, \mu(\mathrm{d}\omega),
\end{equation*}
where $\frac{\mathrm{d}\eta}{\mathrm{d}\mu}\colon\Omega\to\mathbb{R}$ is the Radon-Nikodym derivative of $\eta$ with respect to $\mu$. Let us also assume that the receiver's payoff function is such that $u(a,\cdot)\colon\Omega\to\mathbb{R}$ is bounded and measurable for any action $a\in A$.

Under those conditions, we can apply Proposition 1.4.2 in \cite{Dupuis1997} which implies that:
\begin{equation}\label{eqn:optimal_well_being_infinite}
   \ln\left(\int_{\Omega}e^{\rho u(a,\omega)}\,\mu(\mathrm{d}\omega)\right)=\sup_{\eta\in\Delta(\Omega)}\int_{\Omega}\rho u(a,\omega)\,\eta(\mathrm{d}\omega)-C(\eta,\mu).
\end{equation}
for any $\rho\in\mathopen(0,+\infty\mathclose)$. Moreover, letting $\eta_a\in\Delta(\Omega)$ being the probability measure such that $\eta_a$ is absolutely continuous with respect to $\mu$ and that:
\begin{equation}\label{eqn:optimal_beliefs}
    \frac{\mathrm{d}\eta_a}{\mathrm{d}\mu}(\omega)=\frac{e^{\rho u(a,\omega)}}{\displaystyle\int_{\Omega}e^{\rho u(a,\omega)}\,\mu(\mathrm{d}\omega)},
\end{equation}
for any $\omega\in \Omega$, then the supremum in problem \labelcref{eqn:optimal_well_being_infinite} is attained at the unique maximizer given by $\eta_a$ \citep[see again][Proposition 1.4.2]{Dupuis1997}.

Therefore, adapting the definition of the set of wishful-optimal actions $A(\mu,\rho)$ from the main text (see \cref{eqn:wishful_actions}) to this more general setting, we obtain:
\begin{align*}
    A(\mu,\rho)&=\underset{a\in A}{\arg\max} \max_{\eta\in\Delta(\Omega)} \; \int_{\Omega} u(a,\omega) \, \eta(\mathrm{d}\omega) - \frac{1}{\rho} C(\eta,\mu) \\
    &=\underset{a\in A}{\arg\max} \; \frac{1}{\rho}\ln\left(\int_{\Omega}e^{\rho u(a,\omega)}\,\mu(\mathrm{d}\omega)\right)\\
    &=\underset{a\in A}{\arg\max} \; \int_{\Omega}e^{\rho u(a,\omega)}\,\mu(\mathrm{d}\omega),
\end{align*}
for any $\mu\in\Delta(\Omega)$ and any $\rho\in\mathopen(0,+\infty\mathclose)$. 

\section{Proof for \cref{thm:motivatedaction}}\label{secap:motivatedaction_proof}

Let us study the properties of the belief threshold $\mu_{W}$ as a function of $\rho$ and payoffs. First of all, let us define the function
\begin{equation*}
    \mu_{W}(\rho)=\frac{e^{\rho u(0,0)}-e^{\rho u(1,0)}}{e^{\rho u(0,0)}-e^{\rho u(1,0)}+e^{\rho u(1,1)}-e^{\rho u(0,1)}}.
\end{equation*}
for any $\rho\in\mathopen(0,+\infty\mathclose)$. To avoid notational burden, we omit the subscript $W$ in the proof. We can find the limit of $\mu(\rho)$ at 0 by applying l'Hôpital's rule
\begin{align*}
    \lim_{\rho\rightarrow0}\mu(\rho)&=\lim_{\rho\rightarrow0}\frac{u(0,0)e^{\rho u(0,0)}-u(1,0)e^{\rho u(1,0)}}{u(0,0)e^{\rho u(0,0)}-u(1,0)e^{\rho u(1,0)}+u(1,1)e^{\rho u(1,1)}-u(0,1)e^{\rho u(0,1)}}\\
    &=\frac{u(0,0)-u(1,0)}{u(0,0)-u(1,0)+u(1,1)-u(0,1)}\\
    &=\mu_{B}.
\end{align*}
So, we are back to the case of a Bayesian Receiver whenever the cost of distortion becomes infinitely high. After multiplying by $e^{-\rho u(0,0)}$ at the numerator and the denominator of $\mu(\rho)$ we get
\begin{equation*}
    \mu(\rho)=\frac{1-e^{\rho( u(1,0)-u(0,0))}}{1-e^{\rho( u(1,0)-u(0,0))}+e^{\rho( u(1,1)-u(0,0))}-e^{\rho(u(0,1)-u(0,0))}}.
\end{equation*}
So the limit of $\mu(\rho)$ at infinity only depends on the sign of $u(1,1)-u(0,0)$ as, by assumption, $u(1,0)-u(0,0)<0$ and $u(0,1)-u(0,0)<0$. Hence, $\lim_{\rho\rightarrow+\infty}\mu(\rho)=1$ when $u(1,1)-u(0,0)<0$ and $\lim_{\rho\rightarrow+\infty}\mu(\rho)=0$ when $u(1,1)-u(0,0)>0$. Finally, in the case where $u(0,0)=u(1,1)$ we have
\begin{align*}
    \lim_{\rho\rightarrow+\infty}\mu(\rho)&=\lim_{\rho\rightarrow+\infty}\frac{1-e^{\rho( u(1,0)-u(0,0))}}{2-e^{\rho( u(1,0)-u(0,0))}-e^{\rho(u(0,1)-u(0,0))}}\\
    &=\frac{1}{2}.
\end{align*}
Let us now check the variations of the function. After differentiating with respect to $\rho$ and rearranging terms, one can remark that the derivative of $\mu(\rho)$ must verify the following logistic differential equation with varying coefficient
\begin{equation*}
    \mu'(\rho)=\xi(\rho)\mu(\rho)(1-\mu(\rho)),
\end{equation*}
where
\begin{equation*}
    \xi(\rho)=\frac{u(0,0)e^{\rho u(0,0)}-u(1,0)e^{\rho u(1,0)}}{e^{\rho u(0,0)}-e^{\rho u(1,0)}}-\frac{u(1,1)e^{\rho u(1,1)}-u(0,1)e^{\rho u(0,1)}}{e^{\rho u(1,1)}-e^{\rho u(0,1)}},
\end{equation*}
for all $\rho\in\mathopen(0,+\infty\mathclose)$, together with the initial condition $\mu(0)=\mu_{B}$. Hence, $\xi$ completely dictates the variations of $\mu(\rho)$. Let us study the properties of the function $\xi$ defined on $\mathopen(0,+\infty\mathclose)$. First, still applying again l'Hôpital's rule, its limits are given by
\begin{equation*}
    \lim_{\rho\rightarrow 0}\xi(\rho)=\frac{1}{2}\Bigl(\bigl(u(0,0)-u(0,1)\bigr)-\bigl(u(1,1)-u(1,0)\bigr)\Bigr),
\end{equation*}
and
\begin{equation*}
    \lim_{\rho\rightarrow+\infty}\xi(\rho)=u(0,0)-u(1,1),
\end{equation*}
Second, after rearranging terms, its derivative is given by
\begin{equation*}
    \xi'(\rho)=\frac{\bigl(u(0,0)-u(1,0)\bigr)^2}{\cosh\Bigl(\rho\bigl(u(0,0)-u(1,0)\bigr)\Bigr)-1}-\frac{\bigl(u(1,1)-u(0,1)\bigl)^2}{\cosh\Bigl(\rho\bigl(u(1,1)-u(0,1)\bigr)\Bigr)-1},
\end{equation*}
for any $\rho\in\mathopen(0,+\infty\mathclose)$, where $\cosh$ is the hyperbolic cosine function defined by
\begin{equation*}
    \cosh(x)=\frac{e^{x}+e^{-x}}{2},
\end{equation*}
for any $x\in\mathbb{R}$. Remark that the function defined by
\begin{equation*}
    g(x)=\frac{x^2}{\cosh(\rho x)-1},
\end{equation*}
is strictly decreasing on $\mathopen(0,+\infty\mathclose)$ for any$\rho\in\mathopen(0,+\infty\mathclose)$. So, we have $\xi'(\rho)<0$ and therefore $\mu(\rho)$ strictly decreasing for all $\rho\in\mathopen(0,+\infty\mathclose)$ if and only if $u(0,0)-u(1,0)>u(1,1)-u(0,1)$. Accordingly, $\xi$ is always a strictly monotonic function if and only if $u(0,0)\neq u(1,1)$ and $u(0,1)\neq u(1,0)$. Hence, excluding the extreme case where $u(0,0)=u(1,1)$ and $u(0,1)=u(1,0)$ so $\xi'(\rho)=0$ and $\mu(\rho)=\mu_{B}$ for all $\rho\in\mathopen(0,+\infty\mathclose)$, three interesting cases arise:
\begin{enumerate}[(i)]
    \item If $u(0,0)<u(1,1)$, function $\xi$ has a constant sign for any $\rho\in\mathopen(0,+\infty\mathclose)$ if and only if $u(0,0)-u(0,1)<u(1,1)-u(1,0)$, in which case $\mu_{W}$ is strictly decreasing from $\mu_{B}$ to $0$. In case $u(0,0)-u(0,1)>u(1,1)-u(1,0)$, $\xi$ has a varying sign so $\mu_{W}$ starts from $\mu_{B}$ and is sequentially strictly increasing and strictly decreasing toward $0$.
    \item If $u(0,0)=u(1,1)$, function $\xi$ has a constant sign for any $\rho\in\mathopen(0,+\infty\mathclose)$. In this case $\mu_{W}$ is strictly increasing from $\mu_{B}$ to $1/2$ if and only if $u(0,0)-u(0,1)>u(1,1)-u(1,0)$.
    \item If $u(0,0)>u(1,1)$, function $\xi$ has a constant sign for any $\rho\in\mathopen(0,+\infty\mathclose)$ if and only if $u(0,0)-u(0,1)>u(1,1)-u(1,0)$, in which case $\mu_{W}$ is strictly increasing from $\mu_{B}$ to $1$. In case $u(0,0)-u(0,1)<u(1,1)-u(1,0)$, $\xi$ has a varying sign so $\mu_{W}$ starts from $\mu_{B}$ and is sequentially strictly decreasing and strictly increasing toward $1$.
\end{enumerate}
Accordingly, in case $\mu_{W}$ is non-monotonic in $\rho$, there always exists a unique $\bar{\rho}>0$ such that $\mu_{W}(\bar{\rho})=\mu_{B}$. This concludes the proof.

\section{Proof of \Cref{thm:opt_info}} \label{secap:proof_opt_info}

\subsection{Preliminaries}

In the two next sections, we first prove some properties of the belief indifference cutoff and we then introduce some useful definitions. 

\subsubsection{Properties of the indifference cutoff}\label{secap:proof_indif_cutoff}

In this section, we prove all the relevant properties of the belief indifference cutoff that are needed in the proof of \cref{thm:opt_info}. All the properties that are proven for $\mu_{W}$ are therefore satisfied symmetrically for its inverse bijection $\vartheta$.

\paragraph{Monotonicity}

We first prove that the function $\theta\mapsto\mu_{W}(\rho,\gamma,\theta)$ is strictly increasing for any $\rho\in\mathopen(0,+\infty)$ and $\gamma\in\{0,1\}$. 

When $\gamma=0$, the indifference cutoff is given by
\begin{equation*}
    \mu_{W}(\rho,0,\theta)=\frac{e^{\rho\theta}-1}{e^{\rho}-1}\eqqcolon \mu_{0}(\rho,\theta),
\end{equation*}
for all $\rho\in\mathopen(0,+\infty\mathclose)$ and $\theta\in\mathopen[0,1\mathclose]$. We thus have
\begin{equation*}
     \frac{\partial \mu_{0}}{\partial\theta}(\rho,\theta)=\frac{\rho e^{\rho\theta}}{e^{\rho}-1}>0,
\end{equation*}
for all $\rho\in\mathopen(0,+\infty\mathclose)$ and $\theta\in\mathopen[0,1\mathclose]$, proving the monotonicity of $\mu_{W}(\rho,0,\cdot)$. 

When $\gamma=1$, the indifference cutoff is given by
\begin{equation*}
    \mu_{W}(\rho,1,\theta)=\frac{e^{\rho\theta}-1}{e^{\rho\theta}+e^{\rho(1-\theta)}-2}\eqqcolon \mu_{1}(\rho,\theta),
\end{equation*}
for all $\rho\in\mathopen(0,+\infty\mathclose)$ and $\theta\in\mathopen[0,1\mathclose]$. Let us define the auxiliary function
\begin{equation}\label{eqn:auxiliary}
    \alpha(x)=\frac{x-1}{x+e^{\rho}x^{-1}-2},
\end{equation}
for all $x\in\mathopen[1,e^{\rho}\mathclose]$. First, remark that
\begin{equation*}
    \alpha'(x)=\frac{e^{\rho}(2x-1)-x^2}{(x^2-2x+e^{\rho})^2},
\end{equation*}
for all $x\in\mathopen[1,e^{\rho}\mathclose]$. Therefore, $\alpha'>0$ if and only if $e^{\rho}(2x-1)>x^2$. But, for any $\rho\in\mathopen(0,+\infty\mathclose)$, the quadratic polynomial $x\mapsto e^{\rho}(2x-1)-x^2$ is concave and its roots are given by $x^{-}(\rho)=e^{\rho}-\sqrt{e^{2\rho}-e^{\rho}}<1$ and $x^{+}(\rho)=e^{\rho}+\sqrt{e^{2\rho}-e^{\rho}}>e^{\rho}$. Therefore, $\alpha'(x)>0$ for all $x\in\mathopen[1,e^{\rho}\mathclose]$. Second, we have $\mu_{1}(\rho,\theta)=\alpha(e^{\rho\theta})$. Hence, we have
\begin{equation*}
    \frac{\partial\mu_{1}}{\partial\theta}(\rho,\theta)=\rho e^{\rho\theta}\alpha'\bigl(e^{\rho\theta}\bigr)>0,
\end{equation*}
for any $\rho\in\mathopen(0,+\infty\mathclose)$ and $\theta\in\mathopen[0,1\mathclose]$, proving the monotonicity of $\mu_{W}(\rho,1,\cdot)$. 

\paragraph{Convexity}

Now, we prove that $\theta\mapsto\mu_{W}(\rho,0,\theta)$ is strictly convex for any $\rho$ and that $\theta\mapsto\mu_{W}(\rho,1,\theta)$ admits a unique inflexion point $\theta^{*}$ such that it is strictly convex on the interval $\mathopen[0,\theta^{*}\mathclose)$ and strictly concave on the interval $\mathopen(\theta^{*},1\mathclose]$.

When $\gamma=0$, we have 
\begin{equation*}
     \frac{\partial^2 \mu_{0}}{{\partial\theta}^2}(\rho,\theta)=\frac{\rho^2 e^{\rho\theta}}{e^{\rho}-1}>0,
\end{equation*}
for all $\rho\in\mathopen(0,+\infty\mathclose)$ and $\theta\in\mathopen[0,1\mathclose]$, proving the strict convexity of $\mu_{W}(\rho,0,\cdot)$.

Assume now that $\gamma=1$. Defining the function $\alpha$ as in \cref{eqn:auxiliary} we have
\begin{equation*}
    \frac{\partial^{2} \mu_{1}}{{\partial\theta}^2}(\rho,\theta)=\rho^2 e^{\rho\theta} \alpha'(e^{\rho\theta})+\rho^2 e^{2\rho\theta} \alpha''(e^{\rho\theta}),
\end{equation*}
for all $\rho\in\mathopen(0,+\infty\mathclose)$ and $\theta\in\mathopen[0,1\mathclose]$. Operating the change of variables $x=e^{\rho\theta}$, we can remark that it has the same sign than the expression
\begin{equation*}
    \frac{1}{x}+\frac{\alpha''(x)}{\alpha'(x)},
\end{equation*}
for any $x\in\mathopen[1,e^{\rho}\mathclose]$. Calculations show that
\begin{equation*}
    \frac{\alpha''(x)}{\alpha'(x)}=\frac{2 \big(x^3 - e^{\rho} (3x^2 - 3x + 2) + e^{2\rho}\big)}{\big(x(2-x)-e^{\rho}\big) \big(x^2 - e^{\rho} (2x-1)\big)},
\end{equation*}
for any $x\in\mathopen[1,e^{\rho}\mathclose]$. The goal is thus to solve the equation
\begin{equation}\label{eqn:fucking_hard_equation}
    \frac{2 \big(x^3 - e^{\rho} (3x^2 - 3x + 2) + e^{2\rho}\big)}{\big(x(2-x)-e^{\rho}\big) \big(x^2 - e^{\rho} (2x-1)\big)}=-\frac{1}{x},
\end{equation}
for $x\in\mathopen[1,e^{\rho}\mathclose]$. Solving \cref{eqn:fucking_hard_equation} is equivalent to searching for the solutions of the quartic equation given by
\begin{equation}\label{eqn:quartic}
    x^4 + \big(2-4 e^{\rho}\big) x^3 + \big(4 e^{2\rho} - 2e^{\rho}\big) x -2 e^{2\rho} = 0,
\end{equation}
for $x\in\mathopen[1,e^{\rho}\mathclose]$. \cref{eqn:quartic} factors into
\begin{equation}\label{eqn:quartic_factored}
    \big(x^2 - e^{\rho}\big) \big(x^2 + (4e^{\rho}-2) x -e^{\rho}\big) = 0.
\end{equation}
for any $x\in\mathopen[1,e^{\rho}\mathclose]$. The second factor in \cref{eqn:quartic_factored} is a convex quadratic form which does not admit any root in the interval $\mathopen[1,e^{\rho}\mathclose]$. Indeed, its roots are given by $x^{-}(\rho) = 1 - 2e^{\rho} - \sqrt{1 - 3e^{\rho} + 4e^{2\rho}}<1$ and $x^{+}(\rho) = 1 - 2e^{\rho} + \sqrt{1 - 3e^{\rho} + 4e^{2\rho}}<1$ for any $\rho\in\mathopen(0,+\infty)$. Hence, the only solution of the quartic polynomial belonging to the interval $\mathopen[1,e^{\rho}\mathclose]$ is $x=e^{\rho/2}$. Therefore, for any $\rho\in\mathopen(0,+\infty\mathclose)$, the function $\mu_{1}(\rho,\cdot)$ admits a unique inflexion point at $\theta^{*}=1/2$, and is strictly convex for all $\theta\in\mathopen[0,1/2\mathclose)$ and strictly concave for all $\theta\in\mathopen(1/2,1\mathclose]$.

\paragraph{Asymptotics}

We now turn to the asymptotic properties of the indifference cutoff as a function of $\rho$. Not surprisingly, as $\rho$ vanishes, the cost of belief distortion becomes arbitrarily large, so the indifference cutoff of a wishful agent converges to the indifference cutoff a Bayesian receiver endowed with the same material payoff would have. Indeed, applying L'Hôpital's rule we obtain:
\begin{align*}
    \lim_{\rho\to 0^{+}} \mu_{W}(\rho,\gamma,\theta)&=\lim_{\rho\to 0^{+}} \frac{\gamma\theta e^{\gamma \rho \theta} - (\gamma-1)\theta e^{(\gamma-1) \rho \theta}}{\gamma\theta e^{\gamma \rho \theta} - (\gamma-1)\theta e^{(\gamma-1) \rho \theta} + (1-\theta)e^{\rho (1-\theta)}}, \\
    &=\lim_{\rho\to 0^{+}} \frac{\gamma\theta - (\gamma-1)\theta }{\gamma\theta - (\gamma-1)\theta + (1-\theta)}, \\
    &=\theta,
\end{align*}
for any $(\gamma,\theta)\in T$. Let us now investigate how $\mu_{W}$ converges as $\rho\to+\infty$. First, assume that $\gamma=0$, so we have
\begin{equation*}
    \lim_{\rho\to+\infty}\mu_{0}(\rho,\theta)= \lim_{\rho\to+\infty}\frac{1-e^{-\rho\theta}}{e^{\rho(1-\theta)}-e^{-\rho\theta}},
\end{equation*}
for any $\theta\in\mathopen[0,1\mathclose]$. So, in particular we have
\begin{align*}
    \lim_{\rho\to+\infty}\mu_{0}(\rho,1) &= \lim_{\rho\to+\infty}\frac{1-e^{-\rho}}{1-e^{-\rho}} \\
    &=1.
\end{align*}
and, for any $\theta<1$, we have
\begin{align*}
    \lim_{\rho\to+\infty}\mu_{0}(\rho,\theta) &= \lim_{\rho\to+\infty}\frac{1-e^{-\rho\theta}}{e^{\rho(1-\theta)}-e^{-\rho\theta}} \\
    &=0^{+}.
\end{align*}
Hence, $\mu_{0}(\rho,\cdot)$ converges to the function $\theta\mapsto\mathds{1}_{\theta=1}$ as $\rho\to +\infty$. That is, all receiver types but $\theta=1$ fully motivate action $a=1$, i.e., would opt for the sender's preferred action under any posterior belief.

Now, assume that $\gamma=1$. We have
\begin{equation*}
    \lim_{\rho\to+\infty}\mu_{1}(\rho,\theta)=\lim_{\rho\to+\infty} \frac{1-e^{-\rho\theta}}{1+e^{\rho(1-2\theta)}-2e^{-\rho\theta}},
\end{equation*}
for any $\theta\in\mathopen[0,1\mathclose]$. So, if $\theta<1/2$ then $\lim_{\rho\to+\infty} e^{\rho(1-2\theta)}=+\infty$ and we have
\begin{equation*}
    \lim_{\rho\to+\infty}\mu_{1}(\rho,\theta)=0^{+}.
\end{equation*}
If $\theta=1/2$ then $e^{\rho(1-2\theta)}=1$ and
\begin{align*}
    \lim_{\rho\to+\infty}\mu_{1}(\rho,\theta)&=\lim_{\rho\to+\infty} \frac{1-e^{-\rho\theta}}{2-2e^{-\rho\theta}},\\
    &=\frac{1}{2}.
\end{align*}
Finally, if $\theta>1/2$ then $\lim_{\rho\to+\infty} e^{\rho(1-2\theta)}=+\infty$ so we have
\begin{equation*}
    \lim_{\rho\to+\infty}\mu_{1}(\rho,\theta)=1^{-}.
\end{equation*}
Hence, $\mu_{1}(\rho,\cdot)$ converges to the function $\theta\mapsto \mathds{1}_{\theta\geq 1/2}$ as $\rho\to +\infty$. That is, receiver types such that $\theta<1/2$ fully motivate action 1, i.e., would choose action $a=1$ under any posterior belief unless the sender fully discloses the bad state. Conversely, receiver types such that $\theta>1/2$ fully motivate action 0, i.e., would choose action $a=0$ under any posterior belief unless the sender fully discloses the good state. Finally, receiver type $\theta=1/2$ does not motivate any action and behaves as a Bayesian agent for any $\rho$, as both actions have the same best-case payoff and same payoff variability for him.

\subsubsection{Some useful definitions}

Before delving into the proof, let us introduce some definitions and state (without proof) a useful mathematical result related to real valued functions.
\begin{definition}[S-shapedness]
A function $F\colon \mathopen[0,1\mathclose] \to \mathbb{R}$ is termed (strictly) \emph{S-shaped} if there exists a unique $x^* \in \mathopen(0,1\mathclose)$ such that $F$ is (strictly) convex on the interval  $\mathopen[0,x^*\mathclose)$ and (strictly) concave on the interval $\mathopen(x^*,1\mathclose]$. Conversely, $F$ is (strictly) \emph{inverse-S-shaped} if it is first (strictly) concave and then (strictly) convex.
\end{definition}
\begin{definition}[Single-peakedness]
A function $f\colon\mathopen[0,1\mathclose]\to\mathbb{R}$ is (strictly) \emph{single-peaked} if there exists a unique $x^* \in \mathopen(0,1\mathclose)$ such that $f$ is (strictly) increasing on the interval $\mathopen[0,x^*\mathclose)$ and (strictly) decreasing on the interval  $\mathopen(x^*,1\mathclose]$. Conversely, $f$ is (strictly) \emph{single-dipped} if it is first (strictly) decreasing and then (strictly) increasing.
\end{definition}
\begin{definition}[Single-crossingness]
Let $\bar{\mathbb{R}}=\mathbb{R}\cup\{-\infty,+\infty\}$ denote the extended real line. A function $\varphi\colon \mathopen[0,1\mathclose] \to \bar{\mathbb{R}}$ is (strictly) \emph{single-crossing-from-above} if there exists a unique $x^* \in \mathopen(0,1\mathclose)$ such that $\varphi(x) (<) \leq 0$ for all $x \in \mathopen[0,x^*\mathclose)$ and $\varphi(x) (>) \geq 0$ for all $x \in \mathopen(x^*,1\mathclose]$. Conversely, $\varphi$ is (strictly) \emph{single-crossing-from-below} if it is first (strictly) negative and then (strictly) positive.
\end{definition}
\begin{lemma}\label{thm:charac_S_shaped}
Let $F\colon \mathopen[0,1\mathclose] \to \mathbb{R}$ be a twice-continuously differentiable function and let $f\colon \mathopen[0,1\mathclose] \to \mathbb{R}$ be such that $F'(x)=f(x)$ for all $x\in \mathopen[0,1\mathclose]$. The following conditions are equivalent:
\begin{enumerate}[(i)]
    \item The function $F$ is (strictly) S-shaped (resp.\! inverse-S-shaped).
    \item The function $f$ is (strictly) single-peaked (resp.\! single-dipped).
    \item The function $f'$ is (strictly) single-crossing-from-above (resp.\! from below).
\end{enumerate}
\end{lemma}

\subsection{The proof}

To simplify the presentation of the proof, we first let $\vartheta_{\gamma}(\mu,\theta)=\vartheta(\mu,\gamma,\theta)$ and also let
\begin{equation*}
    \mathcal{A}_{\gamma}(\mu,\rho)=F\bigl(\vartheta_{\gamma}(\mu,\theta)\bigr),
\end{equation*}
denote the expected action of the receiver conditional on $\gamma$.

The regions of convexity and concavity of the sender's indirect utility function are given by the sign of its second order partial derivative with respect to the belief, given by
\begin{equation*}
    \frac{\partial^{2} V}{{\partial\mu}^{2}}(\mu,\rho)=p \frac{\partial^{2}\mathcal{A}_{1}}{{\partial\mu}^{2}}(\mu,\rho) + (1-p) \frac{\partial^{2}\mathcal{A}_{0}}{{\partial\mu}^{2}}(\mu,\rho)
\end{equation*}
for all $\mu\in\mathopen[0,1\mathclose]$ and $\rho\in\mathopen(0,+\infty\mathclose)$, where
\begin{equation*}
    \frac{\partial^{2}\mathcal{A}_{\gamma}}{{\partial\mu}^{2}}(\mu,\rho)=\frac{\partial^{2}\vartheta_{\gamma}}{{\partial\mu}^{2}}(\mu,\rho) f\big(\vartheta_{\gamma}(\mu,\theta)\big) + \left(\frac{\partial\vartheta_{\gamma}}{\partial\mu}(\mu,\rho)\right)^2 f'\big(\vartheta_{\gamma}(\mu,\theta)\big),
\end{equation*}
for all $\mu\in\mathopen[0,1\mathclose]$, $\rho\in\mathopen(0,+\infty\mathclose)$ and $\gamma\in\{0,1\}$.

Let $p\in\mathopen[0,1\mathclose]$. We know from \cref{secap:proof_indif_cutoff} that $\lim_{\rho\to 0^{+}} \mu_{\gamma}(\rho,\theta) = \theta$, which implies that
\begin{equation*}
    \lim_{\rho\to 0^{+}} \vartheta_{\gamma}(\mu,\rho) = \mu
\end{equation*}
for all $\mu\in\mathopen[0,1\mathclose]$ and $\gamma\in\{0,1\}$ and, by continuity of $f$ and $f'$, that
\begin{equation*}
     \lim_{\rho\to 0^{+}} \frac{\partial^{2} V}{{\partial\mu}^{2}}(\mu,\rho) = f'(\mu),
\end{equation*}
for all $\mu\in\mathopen[0,1\mathclose]$. Since $f$ is strictly log-concave, the density function $f$ must be strictly single-peaked on the interval $\mathopen[0,1\mathclose]$, so $f'$ is strictly single-crossing-from above by \cref{thm:charac_S_shaped}. Therefore, since $\rho\mapsto \frac{\partial^{2} V}{{\partial\mu}^{2}}(\mu,\rho)$ is continuous for all $\mu\in\mathopen[0,1\mathclose]$, it must exists an $\varepsilon>0$ such that $\mu\mapsto\frac{\partial^{2} V}{{\partial\mu}^{2}}(\mu,\rho)$ is strictly single-crossing-from-above for all $\rho\in\mathopen(0,\varepsilon\mathclose)$. Letting $\ubar{\rho}=\varepsilon$, this implies that $V(\cdot,\rho)$ is strictly S-shaped for all $\rho\in\mathopen(0,\ubar{\rho}\mathclose]$. Applying Theorem 1 of \cite{Kolotilin2022}, we can conclude that if $\rho\in\mathopen(0,\ubar{\rho}\mathclose]$ then the optimal information policy is upper-censorship for any $p\in\mathopen[0,1\mathclose]$.

Now, assume that $p=0$. Then 
\begin{align*}
    \frac{\partial^{2} V}{{\partial\mu}^{2}}(\mu,\rho)&=\frac{\partial^{2}\mathcal{A}_{0}}{{\partial\mu}^{2}}(\mu,\rho)\\
    &=\frac{\partial^{2}\vartheta_{0}}{{\partial\mu}^{2}}(\mu,\rho) f\big(\vartheta_{0}(\mu,\rho)\big) + \left(\frac{\partial\vartheta_{0}}{\partial\mu}(\mu,\rho)\right)^2 f'\big(\vartheta_{0}(\mu,\rho)\big).
\end{align*}
Thus, in this case, $\frac{\partial^{2} V}{{\partial\mu}^{2}}(\mu,\rho)$ has the same sign than the function $\Phi$ given by
\begin{align*}
    \Phi(\mu,\rho)&=\frac{\displaystyle\frac{\partial^{2}\vartheta_{0}}{{\partial \mu}^2}(\mu,\rho)}{\left(\displaystyle{\frac{\partial\vartheta_{0}}{\partial \mu}(\mu,\rho)}\right)^{2}} \, + \varphi\big(\vartheta_{0}(\mu,\rho)\big),\\
    &=-\rho+\varphi\big(\vartheta_{0}(\mu,\rho)\big)
\end{align*}
for all $\mu\in\mathopen[0,1\mathclose]$ and $\rho\in\mathopen(0,+\infty\mathclose)$, where $\varphi\colon\Theta\to\mathbb{R}$ is the function defined by
\begin{equation*}
    \varphi(\theta)=\frac{f'(\theta)}{f(\theta)}
\end{equation*}
for all $\theta\in\mathopen[0,1\mathclose]$. Since $f$ is strictly log-concave the function $\varphi$ is strictly decreasing and single-crossing-from-above on $\mathopen[0,1\mathclose]$ \citep[see][]{Bagnoli2005}. Moreover, by \cref{secap:proof_indif_cutoff}, the function $\vartheta_{0}(\rho,\cdot)$ is strictly increasing, so the function $\Phi(\cdot,\rho)$ must be strictly decreasing over the interval $\mathopen[0,1\mathclose]$. The function $\Phi(\cdot,\rho)$ is therefore strictly single-crossing-from-above if and only if $\rho<\varphi(0)$, and is non-positive if and only if $\rho\geq \varphi(0)$. Hence, if $\rho<\varphi(0)$ then $\frac{\partial^{2}V}{{\partial\mu}^{2}}(\cdot,\rho)$ is strictly single-crossing-from-above which implies that the sender's indirect utility function $V(\cdot,\rho)$ is strictly S-shaped by \cref{thm:charac_S_shaped}. Theorem 1 of \cite{Kolotilin2022} then implies that the sender's optimal information policy is an upper-censorship policy. Conversely, if $\rho\geq \varphi(0)$ the sender's indirect utility function $V(\cdot,\rho)$ is strictly concave, in which case it is optimal to reveal no information. Letting $\tilde{\rho}=\varphi(0)$ yields the desired result.

Assume now that $p\in\mathopen(0,1\mathclose]$. As we have shown in \cref{secap:proof_indif_cutoff} that $\lim_{\rho\to+\infty} \mu_{0}(\rho,\theta)=\mathds{1}_{\theta=1}$, we must have that
\begin{equation*}
\lim_{\rho\to +\infty} \vartheta_{0}(\mu,\rho) = \lim_{\rho\to+\infty} \frac{1}{\rho}\ln\big(1+(e^{\rho}-1)\mu\big) = \left\{\begin{array}{ll}
   0  & \text{if $\mu=0$} \\
   1  & \text{if $\mu\in\mathopen(0,1\mathclose]$}
\end{array}
\right.,
\end{equation*}
by symmetry with respect to the $45^{\circ}$ degree line. Hence, by continuity of $f$ and $f'$ we must have that
\begin{equation*}
\lim_{\rho\to +\infty} f\big(\vartheta_{0}(\mu,\rho)\big) = \left\{\begin{array}{ll}
   f(0)  & \text{if $\mu=0$} \\
   f(1)  & \text{if $\mu\in\mathopen(0,1\mathclose]$}
\end{array}
\right.,
\end{equation*}
and that
\begin{equation*}
\lim_{\rho\to +\infty} f'\big(\vartheta_{0}(\mu,\rho)\big) = \left\{\begin{array}{ll}
   f'(0)  & \text{if $\mu=0$} \\
   f'(1)  & \text{if $\mu\in\mathopen(0,1\mathclose]$}
\end{array}
\right..
\end{equation*}
Since we also know from \cref{secap:proof_indif_cutoff} that $\lim_{\rho\to +\infty} \vartheta_{1}(\mu,\rho)=\mathds{1}_{\theta\geq 1/2}$, we must have
\begin{align*}
    \lim_{\rho\to +\infty} \vartheta_{1}(\mu,\rho)&=\lim_{\rho\to+\infty}\frac{1}{\rho} \ln\left(\frac{1-2\mu+\sqrt{1+4(e^{\rho}-1)\mu(1-\mu)}}{2(1-\mu)}\right) \\
    &=\left\{\begin{array}{ll}
        0^{+} & \text{if $\mu=0$} \\
        \frac{1}{2} & \text{if $\mu\in\mathopen(0,1\mathclose)$} \\
        1^{-} & \text{if $\mu=1$}
    \end{array}\right..
\end{align*}
Hence, by continuity of $f$ and $f'$ we must have that
\begin{equation*}
\lim_{\rho\to +\infty} f\big(\vartheta_{1}(\mu,\rho)\big) = \left\{\begin{array}{ll}
   f(0)  & \text{if $\mu=0$} \\
   f(1/2)  & \text{if $\mu\in\mathopen(0,1\mathclose)$} \\
   f(1)  & \text{if $\mu=1$}
\end{array}
\right.,
\end{equation*}
and that
\begin{equation*}
\lim_{\rho\to +\infty} f\big(\vartheta_{1}(\mu,\rho)\big) = \left\{\begin{array}{ll}
   f'(0)  & \text{if $\mu=0$} \\
   f'(1/2)  & \text{if $\mu\in\mathopen(0,1\mathclose)$} \\
   f'(1)  & \text{if $\mu=1$}
\end{array}
\right..
\end{equation*}
Meanwhile, we have that
\begin{equation*}
    \frac{\partial\vartheta_{0}}{\partial\mu}(\mu,\rho) = \frac{e^{\rho}-1}{\rho\big(1+(e^{\rho}-1)\mu\big)}
\end{equation*}
that
\begin{equation*}
    \frac{\partial^{2}\vartheta_{0}}{{\partial\mu}^{2}}(\mu,\rho)= \frac{(e^{\rho}-1)^2}{\rho\big(1+(e^{\rho}-1)\mu\big)^{2}}
\end{equation*}
that
\begin{equation*}
    \frac{\partial\vartheta_{1}}{\partial\mu}(\mu,\rho)=\frac{2(e^{\rho}-1)}{\rho\left(1+4(e^{\rho}-1)\mu(1-\mu)+\sqrt{1+4(e^{\rho}-1)\mu(1-\mu)}\right)}
\end{equation*}
and that
\begin{equation*}
   \frac{\partial^{2}\vartheta_{1}}{{\partial\mu}^{2}}(\mu,\rho)=\frac{2(e^{\rho}-1)\left(4(e^{\rho}-1)(1-2\mu)+\frac{4(e^{\rho}-1)(1-2\mu)}{2\sqrt{1+4(e^{\rho}-1)\mu(1-\mu)}}\right)}{\rho\left(1+4(e^{\rho}-1)\mu(1-\mu)+\sqrt{1+4(e^{\rho}-1)\mu(1-\mu)}\right)^{2}}
\end{equation*}

First of all, we have that 
\begin{align*}
    \lim_{\rho\to +\infty} \frac{\partial^{2}\vartheta_{1}}{{\partial\mu}^{2}}(\mu,\rho) &= \lim_{\rho\to +\infty} -\frac{2(e^{\rho}-1)\left(4(e^{\rho}-1)(1-2\mu)+\frac{4(e^{\rho}-1)(1-2\mu)}{2\sqrt{1+4(e^{\rho}-1)\mu(1-\mu)}}\right)}{\rho\left(1+4(e^{\rho}-1)\mu(1-\mu)+\sqrt{1+4(e^{\rho}-1)\mu(1-\mu)}\right)^{2}}\\
    &= \lim_{\rho\to+\infty} (2\mu-1)\frac{1+\frac{1}{2\sqrt{1+4(e^{\rho}-1)\mu(1-\mu)}}}{\rho\left(\frac{1}{e^{\rho}-1}+4\mu(1-\mu)\right)^{2}\left(1+\frac{1}{\sqrt{1+4(e^{\rho}-1)\mu(1-\mu)}}\right)^{2}}\\
    &= \left\{\begin{array}{ll}
        -\infty & \text{if $\mu=0$} \\
        0^{-} & \text{if $\mu\in\mathopen(0,1/2\mathclose)$}  \\
        0 & \text{if $\mu=1/2$}  \\
        0^{+} & \text{if $\mu\in\mathopen(1/2,1\mathclose)$}  \\
        +\infty & \text{if $\mu=1$} 
    \end{array}
    \right..
\end{align*}
In addition, we also have that
\begin{align*}
     \lim_{\rho\to +\infty} \frac{\left(\frac{\partial\vartheta_{1}}{{\partial\mu}}(\mu,\rho)\right)^{2}}{\frac{\partial^{2}\vartheta_{1}}{{\partial\mu}^{2}}(\mu,\rho)}&=\lim_{\rho\to +\infty} -\frac{2(e^{\rho}-1)}{\rho\left(4(e^{\rho}-1)(1-2\mu)+\frac{4(e^{\rho}-1)(1-2\mu)}{2\sqrt{1+4(e^{\rho}-1)\mu(1-\mu)}}\right)}\\
     &=\lim_{\rho\to +\infty} -\frac{2}{4\rho(1-2\mu)\left(1+\frac{1}{2\sqrt{1+4(e^{\rho}-1)\mu(1-\mu)}}\right)}\\
     &=0
\end{align*}
which, since 
\begin{equation*}
    \frac{\partial^{2}\mathcal{A}_{1}}{{\partial\mu}^{2}}(\mu,\rho)=\frac{\partial^{2}\vartheta_{1}}{{\partial\mu}^{2}}(\mu,\rho)\left(f\bigl(\vartheta_{1}(\mu,\rho)\bigr)+\frac{\left(\frac{\partial\vartheta_{1}}{{\partial\mu}}(\mu,\rho)\right)^{2}}{\frac{\partial^{2}\vartheta_{1}}{{\partial\mu}^{2}}(\mu,\rho)} f'\bigl(\vartheta_{1}(\mu,\rho)\bigr)\right),
\end{equation*}
for all $\mu\in\mathopen[0,1\mathclose]$ and $\rho\in\mathopen(0,+\infty\mathclose)$, implies that $\frac{\partial^{2}\mathcal{A}_{1}}{{\partial\mu}^{2}}(\mu,\rho)$ converges to the \emph{same limit} than $\frac{\partial^{2}\vartheta_{1}}{{\partial\mu}^{2}}(\mu,\rho)$ for any $\mu\in\mathopen[0,1\mathclose]$.

Second, we have that 
\begin{align*}
    \lim_{\rho\to +\infty} \frac{\frac{\partial^{2}\vartheta_{0}}{{\partial\mu}^{2}}(\mu,\rho)}{\frac{\partial^{2}\vartheta_{1}}{{\partial\mu}^{2}}(\mu,\rho)} &= \lim_{\rho\to +\infty} \frac{\left(1+4(e^{\rho}-1)\mu(1-\mu)+\sqrt{1+4(e^{\rho}-1)\mu(1-\mu)}\right)^{2}}{\bigl(1+4(e^{\rho}-1)\mu\bigr)\left(4(e^{\rho}-1)(1-2\mu)+\frac{4(e^{\rho}-1)(1-2\mu)}{2\sqrt{1+4(e^{\rho}-1)\mu(1-\mu)}}\right)} \\
    &= \lim_{\rho\to +\infty} \frac{\left(\frac{1}{4(e^{\rho}-1)}+\mu(1-\mu)+\frac{\sqrt{1+4(e^{\rho}-1)\mu(1-\mu)}}{4(e^{\rho}-1)}\right)^{2}}{\left(\frac{1}{16(e^{\rho}-1)^2}+\frac{\mu}{4(e^{\rho}-1)}\right)\left(\frac{1-2\mu}{4(e^{\rho}-1)}\left(1+\frac{1}{\sqrt{1+4(e^{\rho}-1)\mu(1-\mu)}}\right)\right)} \\
    &= 0
\end{align*}
as well as that
\begin{align*}
    \lim_{\rho\to +\infty} \frac{\left(\frac{\partial\vartheta_{0}}{\partial\mu}(\mu,\rho)\right)^{2}}{\frac{\partial^{2}\vartheta_{1}}{{\partial\mu}^{2}}(\mu,\rho)} &= \lim_{\rho\to +\infty} \frac{(e^{\rho}-1)\left(1+4(e^{\rho}-1)\mu(1-\mu)+\sqrt{1+4(e^{\rho}-1)\mu(1-\mu)}\right)^{2}}{\rho\bigl(1+(e^{\rho}-1)\mu\bigr)^{2}\left(4(e^{\rho}-1)(1-2\mu)+\frac{4(e^{\rho}-1)(1-2\mu)}{2\sqrt{1+4(e^{\rho}-1)\mu(1-\mu)}}\right)} \\ 
    &= \lim_{\rho\to +\infty} \frac{\frac{1}{16(e^{\rho}-1)}\left(\frac{1}{4(e^{\rho}-1)}+\mu(1-\mu)+\frac{\sqrt{1+4(e^{\rho}-1)\mu(1-\mu)}}{4(e^{\rho}-1)}\right)^{2}}{\rho\left(\frac{1}{4(e^{\rho}-1)}+\frac{\mu}{4}\right)\left(4(e^{\rho}-1)(1-2\mu)\left(1+\frac{1}{2\sqrt{1+4(e^{\rho}-1)\mu(1-\mu)}}\right)\right)} \\
    &= 0
\end{align*}
which, since $\lim_{\rho\to+\infty}\frac{\partial^{2}\mathcal{A}_{1}}{{\partial\mu}^{2}}(\mu,\rho)=\lim_{\rho\to+\infty}\frac{\partial^{2}\vartheta_{1}}{{\partial\mu}^{2}}(\mu,\rho)$, and
\begin{align*}
    \lim_{\rho\to+\infty} \frac{\frac{\partial^{2}\mathcal{A}_{0}}{{\partial\mu}^{2}}(\mu,\rho)}{\frac{\partial^{2}\vartheta_{1}}{{\partial\mu}^{2}}(\mu,\rho)}&=\lim_{\rho\to+\infty} \frac{\frac{\partial^{2}\vartheta_{0}}{{\partial\mu}^{2}}(\mu,\rho)}{\frac{\partial^{2}\vartheta_{1}}{{\partial\mu}^{2}}(\mu,\rho)} f\bigl(\vartheta_{0}(\mu,\rho)\bigr)+\frac{\left(\frac{\partial\vartheta_{1}}{{\partial\mu}}(\mu,\rho)\right)^{2}}{\frac{\partial^{2}\vartheta_{1}}{{\partial\mu}^{2}}(\mu,\rho)} f'\bigl(\vartheta_{0}(\mu,\rho)\bigr), \\
    &=0,
\end{align*}
implies that $\frac{\partial^{2}\mathcal{A}_{0}}{{\partial\mu}^{2}}(\mu,\rho)$ is negligible compared to $\frac{\partial^{2}\mathcal{A}_{1}}{{\partial\mu}^{2}}(\mu,\rho)$ as $\rho\to +\infty$ for any $\mu\in\mathopen[0,1\mathclose]$.

Therefore, overall, we have that
\begin{align*}
    \lim_{\rho\to +\infty} \frac{\partial^{2}V}{{\partial\mu}^{2}}(\mu,\rho)&= \lim_{\rho\to +\infty} \frac{\partial^{2}\vartheta_{1}}{{\partial\mu}^{2}}(\mu,\rho) \\
    &=\left\{\begin{array}{ll}
        -\infty & \text{if $\mu=0$} \\
        0^{-} & \text{if $\mu\in\mathopen(0,1/2\mathclose)$}  \\
        0 & \text{if $\mu=1/2$}  \\
        0^{+} & \text{if $\mu\in\mathopen(1/2,1\mathclose)$}  \\
        +\infty & \text{if $\mu=1$} 
    \end{array}
    \right..
\end{align*}
which is strictly single-crossing-from below at $\mu=1/2$. Therefore, since $\rho\mapsto \frac{\partial^{2} V}{{\partial\mu}^{2}}(\mu,\rho)$ is continuous for all $\mu\in\mathopen[0,1\mathclose]$, there must exist a $\varepsilon>0$ such that $\mu\mapsto\frac{\partial^{2} V}{{\partial\mu}^{2}}(\mu,\rho)$ is strictly single-crossing-from-below for all $\rho\in\mathopen(\varepsilon,+\infty\mathclose)$. Letting $\bar{\rho}=\varepsilon$, which implies that $V(\cdot,\rho)$ is strictly inverse-S-shaped for all $\rho\in\mathopen(0,\ubar{\rho}\mathclose)$. Applying again Theorem 1 of \cite{Kolotilin2022}, we can conclude that if $\rho\in\mathopen(\bar{\rho},+\infty\mathclose)$ then the optimal information policy is lower-censorship, for all $p\in\mathopen(0,1\mathclose]$.

\end{document}